\documentstyle[psfig]{article}
\font\t=cmbx10 at 16pt
\font\au=cmr10
\font\ad=cmti10 at 9pt
\font\s=cmbx10

\font\fc=cmr10 at 9pt
\def\pmb#1{\setbox0=\hbox{$#1$}%
\kern-.025em\copy0\kern-\wd0
\kern.05em\copy0\kern-\wd0
\kern-.025em\raise.0433em\box0}
\def\bpar {\noindent\hangafter=1\hangindent=0.25 true in}

\begin{document}
\begin{center}
{\t Asymmetries in Powerful Extragalactic Radio Sources
}
\end{center}
\vspace{0.7cm}

\begin{center}
{\au  Gopal-Krishna}\\
{\ad 
National Centre for Radio Astrophysics, Tata Institute of
Fundamental Research, Pune University Campus,  Post Bag No.\ 3, Pune 411 007, India
}\\
\vspace{0.5cm}
{\au Paul J. Wiita}\\
{\ad
Department of Physics \& Astronomy, Georgia State University,
 P.O.\ Box 4106, Atlanta GA 30302-4106, USA
}
\end{center}
\vspace{0.7cm}

\vspace{0.6cm}

\noindent {\it My own suspicion is that the universe is not only queerer 
than we suppose, but queerer than we can suppose} --- J.B.S.\ Haldane

\vspace{0.8cm}











\noindent
{\bf Overview}

\vskip0.2cm

\noindent
This review aims to provide the background and current perspective of 
some of the striking asymmetries exhibited by extragalactic double-lobed 
radio sources, paying special attention to the reported interrelationships 
among some of them.  The asymmetries addressed here are related not only to
the jets, and we particularly aim to highlight some less discussed aspects 
pertaining to the radio lobes and hot-spots for which any relativistic 
beaming effects are milder than those imposed on the jets. Various 
attributes of these features, such as their geometry, intensity, spectrum, 
and polarization provide important underpinnings to models of radio galaxies 
and quasars, as well as to theories that attempt to unify them. These 
constraints can be further sharpened by exploiting the more recently 
discovered asymmetry properties of the extended optical emission-line 
regions (EELR) associated with powerful radio sources, including the giant 
halos of Lyman-$\alpha$ emission associated with them. We suggest that
highly extended disks of dusty gaseous material, which could also be parts 
of cosmic filaments, around the host galaxies of radio sources may hold
the key to understanding some important types of asymmetries, namely, the 
correlated radio--optical asymmetry and perhaps also the depolarization 
asymmetry of the twin radio lobes.

We also discuss the nature and possible implications of the extreme type 
of asymmetry witnessed in the enigmatic quasar 3C273, whose {\it entire} 
radio emission is observed on just one side of the nucleus. In this context, 
theoretical scenarios for the production of one-sided jets are also briefly 
recounted. Taking a clue from the observed one-sidedness of the extended 
radio jets even in `weak-headed-quasars', we also propose a scenario for 
the late evolutionary stage of double radio sources. We end by providing
a list of key questions relating to radio source asymmetries which still 
require final answers. 
 
\vskip 0.3cm

\noindent 
{\s 1. Background and General Perspective}
\\

\noindent
Although a bifurcated ejection of energy from radio-loud active galactic nuclei
(AGN) in two oppositely directed channels had come to be appreciated 
already by the 1960s from the observed similarity and roughly symmetric 
disposition of the twin radio lobes about the parent galaxy, many different 
forms of departures from this symmetry have been discovered during the 
subsequent three decades. Investigations of the systematics and interrelations 
between the various manifestations of {\it asymmetry} have played a
pivotal role in the development of the present physical framework for 
extragalactic double radio sources, as discussed in the next sections.  
With the widespread application of CCD detectors during the last decade, 
the search for structural asymmetries could be successfully extended to 
much shorter wavebands, namely, the ultra-violet, optical, near-infrared, 
and even X-rays. 
The images obtained through the different electromagnetic windows have 
greatly enlarged our vision of the asymmetries, firstly by revealing new 
types of asymmetries and, furthermore, by exposing trends that signify 
important correlations among different manifestations of the asymmetries. 
This encouraging backdrop has provided a strong impetus for the studies 
that could sharpen the theories of double radio sources by confronting them 
with the rapidly widening domain of the observational findings in this area.

The earliest inkling of structural asymmetry came from the conspicuous
one-sidedness of the optical jet already noted by Curtis (1918) in the 
nearby galaxy M87, which was later demonstrated to be a synchrotron 
source (Baade 1956), and whose apparent asymmetry has since been 
clearly demonstrated in the radio and some other wavebands as well. 
The double-lobed radio structure of such radio galaxies was first 
discovered, unexpectedly, from radio interferometric observations 
of the powerful radio galaxy Cygnus A (Jennison \& Das Gupta 1953), 
which led to the concept of (non-spherical) symmetry of radio sources. 
Thus, in a
sense, the concern over an asymmetric characteristic actually predates 
the discussions about the symmetric attributes of external galaxies
(e.g., the double-lobed radio morphology).

Since asymmetry of jets is found to be a ubiquitous feature of quasars, 
the question of `hiding' the jet feeding the other radio lobe led to
the `relativistic beaming' paradigm for the jets (e.g., Rees et al.\ 1982
and references therein), culminating in the 
development of the unified scheme, according to which quasars and radio
galaxies are the same in kind and differ merely in orientation of the
principal axis from our line-of-sight.  While the relativistic jets are 
highly sensitive to orientation effects, the apparent properties such as
the relative brightness and the locations of the slower moving components, 
including  the radio lobes and hot-spots, can also be measurably 
influenced by orientation effects, on account of light-travel-time 
differences and Doppler boosting. Thus, the asymmetries have been effectively
used to infer the dynamics, orientation, intensity evolution and ages of 
such components, laying the boundary conditions for theoretical models
of them.
Additionally, the redshift dependences of the structural asymmetry and 
distortions of double radio sources have been investigated, in order to
probe the cosmological evolution of the radio source properties and their
environment.

The potential of the radio source asymmetry received a major boost
in 1988, following the discovery of the depolarization asymmetry 
of the radio lobe pair of quasars (\S 3). This so called 
Laing--Garrington effect is generally believed to be also an orientation 
effect, and has been instrumental in clinching the long-standing 
debate in favor of the postulated relativistic bulk motion in the 
{\it kiloparsec-scale} radio jets of very powerful radio sources. 
At the moment, the status and understanding
of the spectral index asymmetry, first reported in 1991 (\S 4), is rather 
confusing. From the available results based on rather small samples, it is still  
unclear to what extent the asymmetry must be intrinsic to the sources.

A new dimension was added to such studies with the discovery of the
tightly correlated nature of the structural asymmetries revealed by 
the nonthermal radio and the thermal optical line emission. Two 
radically different perspectives of this phenomenon are discussed 
in \S 5. Another recently discovered optical asymmetry in high-$z$ radio
sources is related to the giant halos of Ly-$\alpha$ emission associated 
with the radio lobes, which are probably excited by the UV photons from the AGN. 
The apparent morphological asymmetries of the Ly-$\alpha$ halos (together 
with their observed kinematical profiles) can provide a powerful means of 
inferring the orientations of the radio sources, the motions of
the host galaxies, and, in particular, 
of estimating the physical properties of the dusty superdisks which appear 
to be commonly associated with high-$z$ radio galaxies (RGs) (\S\S 5 \& 6). 
Aside from being the tracers of dust in the early universe, these 
superdisks could be instrumental in causing several prominent types of
asymmetry.

Observations of asymmetries have proved to be a strong catalyst for
theoretical studies and numerical simulations concerned with the
production and propagation of the AGN beams, including  models for
ejecting one-sided (alternating) jets. Although not favored currently,
such possibilities must still be kept in sight, and in $\S 7$ we provide
some justification for this cautious approach.  Despite our brief
consideration of intrinsic jet asymmetries, we stress that in this
review we are mainly concerned with the asymmetry of mildly beamed
or unbeamed components (the hot-spots and lobes).  Thus we do not
discuss here the huge number of papers focussed on jet asymmetry. 
\\

\noindent 
{\s 2. The radio lobe-length and flux density asymmetries in  FR II sources}
\\
\\
\noindent {\s 2.1 Historical perspective}
\\

\noindent
Although the term ``classical double radio source'' stems from the
roughly symmetric appearance of the two radio lobes about the parent active
galaxy, it is the prevailing degree of the observed asymmetry that holds the key to
constraining many aspects of their physical models and their unification theories.
The potential of such studies has been appreciated and exploited throughout
much of the last four decades.
Thus, while developing one of the earliest (kinematical) models of double radio
sources, Ryle \& Longair (1967) pointed out that light-travel-time effects
and Doppler boosting due to relativistic velocities of the radio lobes 
can cause an apparent asymmetry of the lengths and flux densities of the twin
lobes, if the source axis happens to be oriented away from the sky plane.
De Young \& Axford (1967) gave an early account of the environmental impact in terms of 
ram pressure confinement of the lobes. Within the framework of a simple
 kinematical model, the apparent lobe-length ratio (the asymmetry parameter,
$Q \equiv D_{\rm approach}/D_{\rm recede}$) and the apparent flux ratio
($R \equiv F_{\rm approach}/F_{\rm recede}$) of the approaching and
receding lobes are (e.g., Komberg 1994; Ryle \& Longair 1967)

\begin{equation}
Q = (1+\beta{\rm cos}\theta)/(1-\beta{\rm cos}\theta),
\end{equation}
and
\begin{equation}
R = [(1+\beta{\rm cos}\theta)/(1-\beta{\rm cos}\theta)]^{n-\alpha}, 
\end{equation}
\noindent with $c\beta$ the bulk velocity of the emitting plasma (assumed for 
simplicity to be the same on both sides),
$\theta$ the angle between their motions and the line-of-sight to the observer,
$\alpha$ the spectral index (defined so $S_{\nu} \propto \nu^{+\alpha}$) and 
$n = 2$ or $3$ depending upon whether emission from a
continuous flow or discrete blobs, respectively, is assumed.  For the hot-spots and
lobes, typical values of the spectral index are $-1.0 < \alpha < -0.5$, though
some high-redshift sources exhibit steeper spectral indices ($\alpha < -1$).

Since the receding lobe is farther from us, the ages (measured in their rest
frames) at which the two lobes are observed are in the ratio
\begin{equation}
t_{\rm recede}/t_{\rm approach} = (1-\beta{\rm cos}\theta)/(1+\beta{\rm cos}\theta). 
\end{equation}

After the discovery of compact hot-spots near the outer edges of
radio lobes (Hargrave \& Ryle 1974; Hargrave \& McEllin 1975), it became clear
that the asymmetry can be more meaningfully exploited using them (instead 
of the entire lobe).  In that the main thrust of the collimated energy supply 
from the galactic nucleus is focussed through the jets into the hot-spots,
any relativistic motion is more likely to be associated with them
(Longair, Ryle \& Scheuer 1973; Blandford \& Rees 1974; Scheuer 1974),
and not with the extended lobes.  However, while the flow down the jets
may well remain highly relativistic all along, with $\beta_{\rm jet} \sim 1
$ (e.g.\ Georganopoulos \& Kazanas 2004; \  Urry \& Padovani 1995), we 
shall see that the hot-spot advance speed, $\beta$, is much slower.

Numerous studies since then have focussed on the use of the core--hot spot
separation ratio as the measure of the `asymmetry parameter', $Q$, beginning with 
the early works by Fomalont (1969) and Ingham \& Morrison (1975).
Assuming an intrinsic source symmetry and a random orientation of the
source axis relative to the line-of-sight, these and other authors were able to
reproduce the observed distribution of $Q$ among 3CR radio sources (having a 
median $Q$ of $\sim~1.3$) with a typical value of $\sim 0.2c-0.3c$ for the 
speeds of the hot-spots (e.g., Longair \& Riley 1979; Banhatti 1980). 
Similar speeds were estimated by Macklin (1981),
employing the $Q$-distribution for a well-defined set of 76 3CR sources, and simultaneously
fitting the observed distribution of the misalignment angle of the two
lobes relative to the nucleus of the parent galaxy. Nonetheless, all 
these authors concurred that such estimates only provide upper bounds to
the true speeds, since part of the apparent asymmetry is 
likely to be intrinsic to the source.  Earlier, Ryle \& Longair (1967) had 
argued that in order for their simple kinematic model, based on the 
inference of relativistic motion of the lobes, to work, the radio luminosity 
of the lobes must rapidly `evolve' with time ($P~\propto~t^{-(3-\alpha)}$). 
Such a steep temporal fading, however, followed from their assumption that 
the lobes of double radio sources are a one-time ejection phenomenon.
 
The first major claim in favor of non-relativistic speeds of the hot-spots 
came from the work of Mackay (1973); he used the observed distribution of
the lobe flux ratios, $R$, for a well-defined subset of the
 3CR complete sample then just mapped with the Cambridge 
1-mile synthesis telescope, to set an upper bound of $0.08c$ to
the speed of most of the lobes. A prediction of the simple kinematic model 
is that the approaching lobe should appear not only more distant from the 
optical identification (due to the light travel time effect), but 
also somewhat brighter, due to Doppler boosting. However, Mackay (1971) found 
the opposite: the brighter of the two lobes was seen closer to the optical 
identification in a vast majority of the 3CR sources. This, so called, 
`Mackay's rule' provided a strong hint that double sources recede from the
nucleus non-relativistically and that their apparent asymmetry must be of
largely intrinsic origin (see, also, Ingham \& Morrison 1975).  
Subsequently, Macklin (1981) was able to demonstrate that usually the 
hot-spots are more asymmetric in brightness than 
are the diffuse tails. He further showed that the Mackay's rule is primarily
applicable to the diffuse structure (rather than to the hot-spots), and that
this could be a selection effect arising from the heavier expansion losses
for the diffuse tail associated with the more distant hot-spot, which 
presumably resides in a less dense environment (see also, Valtonen 1979).  
Indeed, a simple way to understand Mackay's rule would be to imagine that a 
double radio source is born in an external environment having a density 
gradient.  The lobe moving on the denser side would then be shorter due to
larger ram-pressure, but at the same time, brighter,
due to a better confinement (Hintzen \& Scott 1978; Gopal-Krishna \& Wiita
1991; Saikia et al.\ 1995, 2003; Jeyakumar et al.\ 2004). Clearly, in
this picture the lobe-length asymmetry produced by the dominance of
the environmental factor is expected to be weaker for more energetic sources,
which again is consistent with observations (e.g., Zi{\c e}ba \& Chy\.zy 1991;
Saikia et al.\ 1995, 2003).

Currently, the most popular explanation for the intrinsic structural asymmetry 
invokes an unequal density of the circumgalactic medium, on the 
scale of tens of kiloparsecs (e.g., Miley 1980; Swarup \& Banhatti
1981; Ensman \& Ulvestad 1984; Fokker 1986; Liu \& Pooley 1991b; McCarthy, van Breugel \& Kapahi 
1991, hereafter, MvBK; Lister, Hutchings \& Gower 1994, hereafter, LHG).
However, from a new interpretation of the observed correlated radio-optical
asymmetry of radio galaxies (MvBK; Liu, Pooley, \& Riley 1992), it has 
been more recently
inferred that a substantial part of the (intrinsic) asymmetry may be
contributed by the motion of the parent galaxy itself (Gopal-Krishna
\& Wiita 1996, 2000; \S 5.1). In either case, the dominance of intrinsic 
asymmetry ($Q_{\rm int}~\neq~1$) excludes the possibility of applying the
elegant formulation of Banhatti (1980) to determine the 
velocity distribution {\it analytically} from the observed $Q$-distribution. 
It may also be mentioned that several researchers (e.g., Ry\'s 1994; LHG) have 
attempted to model the $Q$-distribution within the framework of the 
alternate ejection hypothesis proposed by Rudnick \& Edgar (1984)
(see also, Willis, Wilson \& Strom 1978; Wiita \& Siah 1981; Komberg 1994), 
though differing conclusions have been reached (\S 7.2).  The first 
treatment of the lobe expansion  which treated the dynamics of a
relativistic hot-spot was given by Gopal-Krishna, Wiita \& Saripalli (1989).
A recent attempt by Arshakian \& Longair (2000) to generalize the $Q$-distribution
analysis of Banhatti (1980) to include both intrinsic and relativistic sources of the
asymmetries concludes that they are actually of comparable importance (also see
\S 2.4) for their entire sample of RGs and quasars, but that, perhaps
unsurprisingly, the relativistic effects are more important for quasars
and the intrinsic ones more so for RGs.
\\

\noindent {\s  2.2.  Subtleties of the asymmetry parameter, Q}
\\ 

\noindent
As mentioned above, using the 3CR complete sample, Macklin (1981) 
demonstrated that Mackay's rule primarily applies to the diffuse tails in the
lobes (rather than to the hot-spots), and that this could well be a selection 
effect. Subsequent studies of the 3CR sample and other samples have led to
interesting findings, which are of considerable import for the theories of
radio source evolution and their interaction with the environment.


\noindent $\bullet$ {(i)   Radio galaxies become more symmetric with
increasing radio luminosity
      (Zi{\c e}ba \& Chy\.zy 1991). However, their asymmetry increases 
      with redshift, $z$ (Zi{\c e}ba \& Chy\.zy 1991; MvBK). The former trend
      would be consistent with the high degree of lobe-length asymmetry
      ($Q_{\rm median}\sim~2)$ observed in giant radio galaxies (GRGs) which 
      have intermediate radio luminosities and for which projection is least 
      likely to be a significant factor (see Saripalli et al.\ 1986; 
      Parma et al.\ 1996; Arshakian \& Longair 2000;
Ishwara-Chandra \& Saikia 2002).  It appears that $Q$, the overall projected 
linear sizes ($\ell$),  and the apparent 
bending angles all evolve with $z$ more strongly for RGs than
for quasars (Chy\.zy \& Zi{\c e}ba 1995).

\noindent $\bullet$ {(ii)  An early inkling of quasars not following
Mackay's rule came 
      from the work of Riley \& Jenkins (1977) on 3CR 
      quasars. More specifically, it was demonstrated by Teerikorpi
      (1984, 1986) that among quasars, Mackay type asymmetry prevails 
      only up to a certain luminosity, beyond which the sense of asymmetry 
      is reversed. The critical luminosity is $\approx
      10^{27}$W Hz$^{-1}$ at 1 GHz, taking $H_0~=~75~$km s$^{-1}$ Mpc$^{-1}$.
      and $q_o~=~0.5$. His analysis favored the possibility that high-luminosity 
      quasars (rather than high$-z$ quasars) tend to disobey Mackay's rule.
      With  hindsight, this is what would be expected, since it is in
      more luminous quasars that the light-travel-time effects are expected
      to be large, due to a combination of their hot-spot moving with a 
      higher speed and in a direction relatively close to the line-of-sight 
      (within the framework of the `unified scheme'; \S 2.3). This point is 
      further discussed below.}
      
\noindent $\bullet$ {(iii) For quasars, there seems to be no strong
dependence of $Q$ on redshift,
       at least up to $z \sim~2$ (e.g., Kapahi 1990a; also, LHG). 
       In contrast, for 3CR RGs, MvBK have noticed a statistical tendency for $Q$ to
increase with redshift, over the range $0 < z < 2.5$, as also suggested
by Zi{\c e}ba \& Chy\.zy (1991). Regardless of whether the lobe-length 
asymmetry arises from environmental asymmetry/inhomogeneity (\S 5), or 
quite substantially from the motion of the parent galaxy itself (\S 5.1; 
Gopal-Krishna \& Wiita 1996), the increased asymmetry found for galaxies at high 
redshifts would be expected to be manifested through a corresponding rise in the apparent 
bending angle, $\Psi$, (a measure of departure from colinearity of the twin 
lobes with the nucleus), particularly since at higher redshifts the physical 
sizes of radio sources were smaller (Kapahi 1989; Singal 1988). 
Indeed, a comparison of radio structures of  lobe-dominated 
quasars at $z \le~1.5$ and $z \ge~1.5$ led Barthel \& Miley (1988) to suggest 
that quasars are more bent at higher redshifts. However, this oft-cited 
claim was questioned by Kapahi (1990b) 
who failed to find a strong effect, in spite of using maps that were better 
matched in {\it linear} resolution. Settling this issue is of considerable 
significance for cosmology. Probably a more appropriate approach would be to
employ samples of RGs alone (as also used for investigating the $Q - z$
dependence), because their structures are believed to be less distorted by 
projection effects, and they are now being found in substantial numbers
at high redshifts. 
\\

\noindent{\s 2.3 Lobe-length asymmetry in the context of the Unified Scheme for 
powerful radio sources}
\\

\noindent
According to a widely discussed unified scheme for powerful radio galaxies
and quasars, the latter are RGs with the AGN axis oriented within 
a certain critical angle ($\theta_c \approx 45^{o}$) from the line-of-sight 
(Barthel 1989; also, Scheuer 1987; Peacock 1987; Morisawa \& Takahara 1987; 
Antonucci \& Barvainis 1990). 
In an early examination of orientation effects, a correlation between the 
two orientation dependent parameters, $Q$ and the core-fraction, $f_c$ 
(the flux ratio of the radio core flux to total flux) was searched for 
using a set of 26 3CR sources with detected radio cores (Gopal-Krishna 1980). 
Subsequently, using larger samples of quasars, a weak but significant 
$Q$--$f_c$ correlation was found (e.g., Kapahi \& Saikia 1982;   Saikia 1984;
Hough \& Readhead 1989), though the 
effect has not been discerned in some other quasar samples (e.g., Lister, Gower
\& Hutchings  1994). Baryshev \& Teerikorpi (1995) 
have analyzed the asymmetry 
properties of double RGs within the context of the unified
scheme.  They allowed for intrinsic asymmetries, including
non-zero misalignment angles, and performed numerical studies to bolster
their analytical models for velocities and other properties of quasars
and RGs.
More recently, Saikia et al.\ (1995, 2003)
 have found 
evidence for a higher degree of asymmetry among compact-steep-spectrum (CSS) 
quasars, {\it vis-\`a-vis} CSS RGs, again in broad agreement with the 
orientation-based unified scheme. Such asymmetries are
reproducible 
through propagation of symmetric jets through asymmetric confining media  
as shown by the analytical
models and numerical simulations of Jeyakumar et al.\ (2004). 

One important check on the consistency of the $Q$-distributions with the 
unified scheme comes from the work of Best et al.\ (1995),
which employs the well observed 3CRR complete sample. For a subset of 95
FR II sources with a well-identified hot spot in each lobe, they found that
the quasars appear distinctly more asymmetric than the radio galaxies and
that this difference is not due to any luminosity or redshift bias.
Likewise, they also found that, compared to galaxies, the radio lobe-pair in
quasars appear more misaligned (i.e., less colinear with the nucleus). 
Note that, unlike $Q$, the apparent misalignment angle $(\Psi$) of the 
twin-lobes 
does not depend upon the hot-spot speed. Taking the critical viewing angle 
($\theta_c$), which separates galaxies from quasars in the unified scheme,
to be $45^o$ (Barthel 1989),
Best et al.\ (1995) deduced that typically the {\it intrinsic} misalignment angle of 
the hot-spots is randomly distributed within a cone angle $\phi_{max}~=~
10^o$. Adopting these values of $\phi_{max}$ and $\theta_c$, they could then
reproduce the observed marked difference between the $Q$-distributions for the 
RGs and quasars, taking typical hot-spot speeds of 0.2$c$ -- 0.3$c$. 
Further, making an allowance for an (environmentally induced) intrinsic
asymmetry, they deduced a marginally lower value of $\sim 0.16c$. Similar 
estimates had earlier been reported by LHG, from their orientation-based 
modeling of the radio structural parameters for a large (but not 
complete) sample of 192 powerful RGs and quasars. 
 
While such large hot-spot speeds conform to the estimates coming from
the analysis of synchrotron ageing effects in the radio lobes in powerful 
sources at $z~\ge~0.2$ (e.g., Liu, Pooley \& Riley 1992; also, Leahy,
Muxlow \& Stephens 1989),  
an analysis of the $Q$-distributions of quasars by Scheuer (1995), 
within the framework of the unified theory (which imposes a $\theta_c$), has
indicated that even for the most powerful sources, the hot-spot speeds are
only a few percent of the speed of light.  As discussed below (\S 2.4), the extra clue 
employed by Scheuer is the apparent jet-sidedness, which can be used to identify the 
approaching and receding lobes (see, also, Saikia 1984). 
Thus, the origin of the substantially higher values of $\beta$ 
obtained in the analysis of Best et al.\ (1995) remain to be fully understood, 
particularly in view of their assertion that such high speeds are necessary to
explain the observed contrast between the distributions of $Q$ and $\Psi$ 
for powerful RGs and quasars, within the framework of the unified 
scheme. 

	The differences in cosmological evolution of $\ell$ between
RGs and quasars, as well as the stronger correlations of
$\ell$ with radio luminosity for RGs than for
quasars, may imply difficulties for the pure orientation based unified scheme (Chy\.zy \&
Zi{\c e}ba 1995).  We note, however, that these differences can be readily understood 
within a more
realistic unified scheme which takes into account the growth of radio sources
with time as well 
as a rise in $\theta_c$ with intrinsic source
power (Gopal-Krishna, Kulkarni \& Wiita 1996; also, Lawrence 1991).
\\

\noindent{\s 2.4 Further constraining the hot-spot speeds by the jet-sidedness}
\\


\noindent
As mentioned above, Scheuer (1995) has argued that a much sharper 
interpretation of the $Q$-distribution is possible by invoking additional 
information, namely, the presence of the jet, which allows one to distinguish
the approaching lobe/hot-spot from its receding counterpart. This clue
was earlier employed by Saikia (1984), but his sample of 36 quasars did not
show the expected trend; namely a statistical excess of the lobe-length
on the jet side. Since the orientation dominated effects are likely to be
stronger for highly luminous radio sources with their hot-spots advancing 
more rapidly, Scheuer has determined the lobe-length ratios for 3 essentially 
independent samples containing a total of 
43 sources with conspicuous jets (including 41 quasars), above an emitted
integrated luminosity of $\sim 10^{27.5}$W Hz$^{-1}$sr$^{-1}$ at 178 MHz.
For each of these high luminosity samples, he finds the jet-side lobe 
to be systematically, albeit only marginally, longer than the opposite 
(receding) lobe. This is basically in agreement with the kinematical model.
Assuming that $\theta$ is uniformly distributed over solid angle between
$\theta~=~0$ and $\theta~=~\theta_{\rm max}$, the predicted median of $Q$ 
and the mean of the logarithm of $Q$ are (Scheuer 1995):
\begin{equation}
Q_{\rm median} = {1 + 0.5 \beta(1 + {\rm cos}\theta_{\rm max}) \over 
{1 - 0.5 \beta(1 - {\rm cos}\theta_{\rm max})}}
= \beta(1 + {\rm cos}\theta_{\rm max})[1 + {1 \over 3}\beta^2(1+{\rm
cos}\theta_{\rm max})^2 \dots], 
\end{equation}
\begin{equation}
\langle{\rm ln}Q\rangle = \int_{\theta=0}^{\theta_{\rm max}} {\rm ln}Q~ {\rm d}P
= {f(\beta) - f(\beta {\rm cos}\theta_{\rm max}) \over \beta(1 - {\rm
cos}\theta_{\rm max})}, 
\end{equation}
\noindent where $f(\beta) = (1+\beta)~{\rm ln}(1+\beta) +
(1-\beta)~{\rm ln}(1-\beta)$.

Comparison of the observed $Q$-distributions with the predicted ones yields
the fairly robust results: (i) $\beta \ge~0$, and (ii) $\beta  \le~0.15$, with the 
most probable value being $\beta = 0.03~\pm~0.02$ (Scheuer 1995). It is 
remarkable that these estimates of $\beta$ from the source asymmetry 
(which, obviously, are independent 
of the adopted values of the cosmological parameters $H_0$, $\Omega$ and 
$\Lambda$) are well below the corresponding value of $\beta~\sim~0.165$ deduced
by Scheuer from the synchrotron ageing results reported by Liu, Pooley 
\& Riley (1992). The substantial difference between the two estimates was
taken by Scheuer as
evidence for a rapid `backflow' of the
relativistic plasma within the lobes, which is indeed expected if the jet material 
is much less dense than the ambient plasma.  Such strong backflows
are clearly observed in many numerical simulations of jet
propagation (e.g., Norman 1996; Hooda \& Wiita 1998). 

It is worth stressing that
even though the synchrotron
ageing estimates are still routinely used, they 
have been criticized from several other different
angles (e.g.\ Wiita \& Gopal-Krishna 1990; Siah \& Wiita 1990; 
Rudnick, Katz-Stone \& Anderson 1994; Eilek \& Arendt 1996;
Blundell \& Rawlings 2000).
See \S 5.2 for a more detailed discussion of radio galaxy ages.

The most recent  analysis of high quality  data leading to a
determination of $\beta$ for the hot-spots is due to Arshakian \& Longair (2000).
They also use the jet-sidedness ratio, which they could estimate for
80 per cent of the FR II sources in the 3CRR complete sample, giving them
a total of 103 sources (71 RGs and 32 quasars).  They conclude that the 
asymmetry is dominated by relativistic effects for
the quasars, in the sense that the approaching jet is typically on the
long lobe side.  But just the opposite is the case for RGs,
where there are more sources with approaching jets on the short lobe
side; this is indicative of the intrinsic asymmetry (or host motion
asymmetry) for these objects.  Their best estimate of the mean hot-spot
velocities is $\langle\beta\rangle = 0.11~\pm~0.013$, substantially above Scheuer's,
and not too dissimilar from the results estimated using the
unreliable spectral ageing technique.
Within the framework of a Gaussian model, the standard deviation
of the velocity distribution is found to be  $\sigma_\beta = 0.04$.
Arshakian \& Longair (2000) note that these results are in agreement
with the unified model with $\theta_c \simeq 45^{\circ}$.
Although this work is rather convincing, it is clear that substantially 
larger samples of RGs and quasars with well defined jet-sidedness
will be needed to finally settle the question of the average
value of $\beta$ and if it evolves with $P$, $z$, and/or $\ell$. 
\\

\noindent{\s 2.5. Brightness asymmetry of hot-spots}
\\

\noindent
Besides the asymmetric locations of the hot-spots relative to the core
(\S\S 2.1--2.2), another striking asymmetry related to the hot-spots has
recently been noticed. This new result concerns the brightness asymmetry of the 
hot-spots, which is specially pronounced in the X-ray emission from the
hot-spots of quasars and broad line radio galaxies. 

Although just a handful of such examples 
are known at present, the trend seems quite clear that the brighter of 
the two hot-spots in such sources lies on the side of the main jet
(Georganpoulos \& Kazanas 2003).  These
authors argue that this X-ray asymmetry arises from the inverse Compton boosting
of the radio photons produced inside the hotspot, where the jet plasma undergoes
a rapid deceleration, resulting in the emission of inverse Compton 
X-rays which is beamed more narrowly than is the nonthermal radiation at lower 
frequencies. The hotspot asymmetry caused by such a energy dependent 
beaming furnishes an important piece of evidence for the idea that bulk motion 
of the jet remains relativistic all the way to the hot-spots 
(Georganpoulos \& Kazanas 2003). With the increasing detections of 
hot-spots with {\sl Chandra}, rapid progress in this area can be expected.  
\\

\noindent 
{\s 3. The asymmetric depolarization of the twin radio lobes}
\\


\noindent
The anomalously large difference between the Faraday rotation measure of
the radio lobe pair was first discussed for the case of Cygnus A, and was
interpreted in terms of differential Faraday depth to the two lobes
(Slysh 1966). In 1988, a striking correlation, now 
called the `Laing-Garrington (L-G) effect' was demonstrated using a sample of
double-lobed radio sources (mostly quasars) showing one-sided jets. According
to the correlation, the lobe on the jet-side depolarizes less rapidly with 
increasing
wavelength than does the opposite lobe (Laing 1988; Garrington et al.\ 1988). 

The currently popular explanation of this L-G effect, advanced by its 
discoverers, posits that distant radio sources 
are embedded within $\approx 100~$kpc diameter halos of hot gas, similar
to the cores of the intra-cluster medium (ICM) associated with nearby clusters 
of galaxies. Consequently, the lobe on the near-side, which is also observed to
be associated with the (presumably Doppler-boosted) extended jet, is viewed 
through a smaller magneto-ionic depth of the halo and hence suffers less 
depolarization (see, also, Tribble 1992). For quasars, whose axes are
believed to be oriented away from the plane of the sky, it is reassuring 
that the depolarization asymmetry is indeed found to be governed by the
jet-sidedness, rather than by any difference in the lobe-lengths (Garrington,
Conway \& Leahy 1991). Furthermore, for RGs, for which
the differential magneto-ionic depth to the two lobes would normally be a 
less crucial factor (e.g., Barthel 1989), it is the shorter lobes which tend to 
suffer greater depolarization. This  can be readily understood as a result 
of radially declining density profile of the halo plasma (Garrington \& 
Conway 1991; Tribble 1992; Pedelty et al.\ 1989; also, Strom \& J\"agers 1988). 

Although the above explanation for the L-G effect in (distant) quasars, 
invoking an ICM-core like ambient medium, has almost become a folklore, a few 
disconcerting aspects have been highlighted by Gopal-Krishna \& Nath 
(1997), leading them to consider an alternative scenario.
Firstly, the viability of the `standard' explanation rests on the tacit 
assumption that in every case, the postulated ICM core around the radio source
is somehow able to maintain a diameter
very close to the steadily growing size of the source. Even a factor
of two discrepancy between the two sizes would seriously erode the correlation
(Garrington \& Conway 1991). However, no regulatory mechanism has been
suggested that would ensure the needed tight coupling between the two sizes.
Secondly, whilst the L-G effect is found to become stronger at higher redshifts,
the density, magnetic field, structure, and even the existence, of a hot ICM
core around all high-$z$ quasars is quite uncertain. In the absence of such
deep, high resolution X-ray observations, the basic assumption of the 
canonical model, namely, the
general existence of a dense ICM core ($n > 10^{-2}$cm$^{-3}$) around $z > 1$ 
quasars (Garrington \& Conway 1991) is fraught with uncertainty (though,
clearly, even distant radio sources must still have a low-density gas envelope).
\\

\noindent{\s 3.1.  An alternative scenario for the origin of depolarization asymmetry}
\\

\noindent
In view of the above caveats, it has been proposed that the required 
Faraday screen contributing the extra rotation measure (RM) 
toward the far-side lobe is associated not with any putative ICM, but, instead,
with a large disk of magneto-ionic medium surrounding the host galaxy
and oriented roughly perpendicular to the radio source axis (Gopal-Krishna
\& Nath 1997; Gopal-Krishna \& Wiita 2000). As discussed below, there is a growing evidence for large disks
of dust and gas around RGs and quasars, even at high redshifts 
where the L-G effect is found to be strong. Such dusty {\it superdisks},
or {\it fat pancakes}
could be responsible for a variety of phenomena associated with radio sources 
(\S 5). Their prominent tracers, on smaller scale, are the extended 
dust-lanes 
which tend to align perpendicular to the jets (Kotanyi \& Ekers 1979). They  
have been detected in roughly half of the bright ellipticals, despite the 
difficulty of detection in the absence of a fortuitously edge-on view
(Goudfrooij \& de Jong 1995 and references therein).
At large redshifts ($z > 2)$, evidence for such dusty {\it superdisks}
around RGs and quasars comes from the apparent suppression of the 
Ly-$\alpha$ emission associated with the far-side lobe of radio galaxies
(Gopal-Krishna et al.\ 1995) and quasars (Heckman et al.\ 1991a) (\S 6). 
According to a widely discussed scenario, dusty disks around the elliptical
hosts of powerful radio sources form as a result of accretion of  
gas-rich galaxies, which also provides a trigger for the nuclear
activity (e.g., Fukugita, Hogan \& Peebles 1996; Lynden-Bell
1996; Wilson \& Colbert 1995; Colina \& de Juan 1995; also, Sofue \& 
Wakamatsu 1992). In fact, optical observations have revealed an abundant
population of gas-rich field galaxies undergoing massive starbursts at 
redshifts down to $z \sim 1$ (e.g. Cowie, Hu \& Songaila 1995).
Alternatively, these superdisks could be parts of the cosmic web of
filamentary network of proto-galactic material, as revealed in a 
number of simulations of the structure formation process (e.g., Cen \&
Ostriker 1999).

Independent evidence for the putative {\it superdisks} around radio galaxies has 
emerged from observations in the radio band (see Gopal-Krishna \& Wiita 2000
for a more comprehensive review). For instance, millimetric continuum emission
believed to arise from an enormously massive dusty disk ($M_{\rm gas}~\sim
10^{11-12}~M_{\odot}$) has been detected in the extremely distant ($z = 3.8$)
radio galaxy, 4C41.17 (Dunlop et al.\ 1994; Chini \& Kr\"ugel 1994).
Striking evidence for fat and highly extended gaseous disks
around the elliptical hosts of RGs comes from the detection of
sharp and often quasi-linear inner boundaries of the radio lobes, creating
strip-like central gaps in the radio bridges (Gopal-Krishna \& Wiita 1996; 2000).
Although their detection requires not only maps with high sensitivity and 
dynamic range, but also a favorable orientation of the lobe axis (i.e., close
to the plane of the sky), clear examples of such sharp inner boundaries of
radio lobes have been observed in a number of RGs. These include:
3C34 (Johnson, Leahy \& Garrington 1995); 3C227 (Black et al.\ 1992), 0828+32 and 
3C382 (Capetti et al. 1993) and 3C16,
3C33, 3C61.1, 3C184.1, 3C341, 3C381, 4C14.11 and 4C14.27 (Leahy \& 
Perley
1991). An example of such morphology is shown in Fig.\ 1.

\begin{figure}
\centerline{\psfig{figure=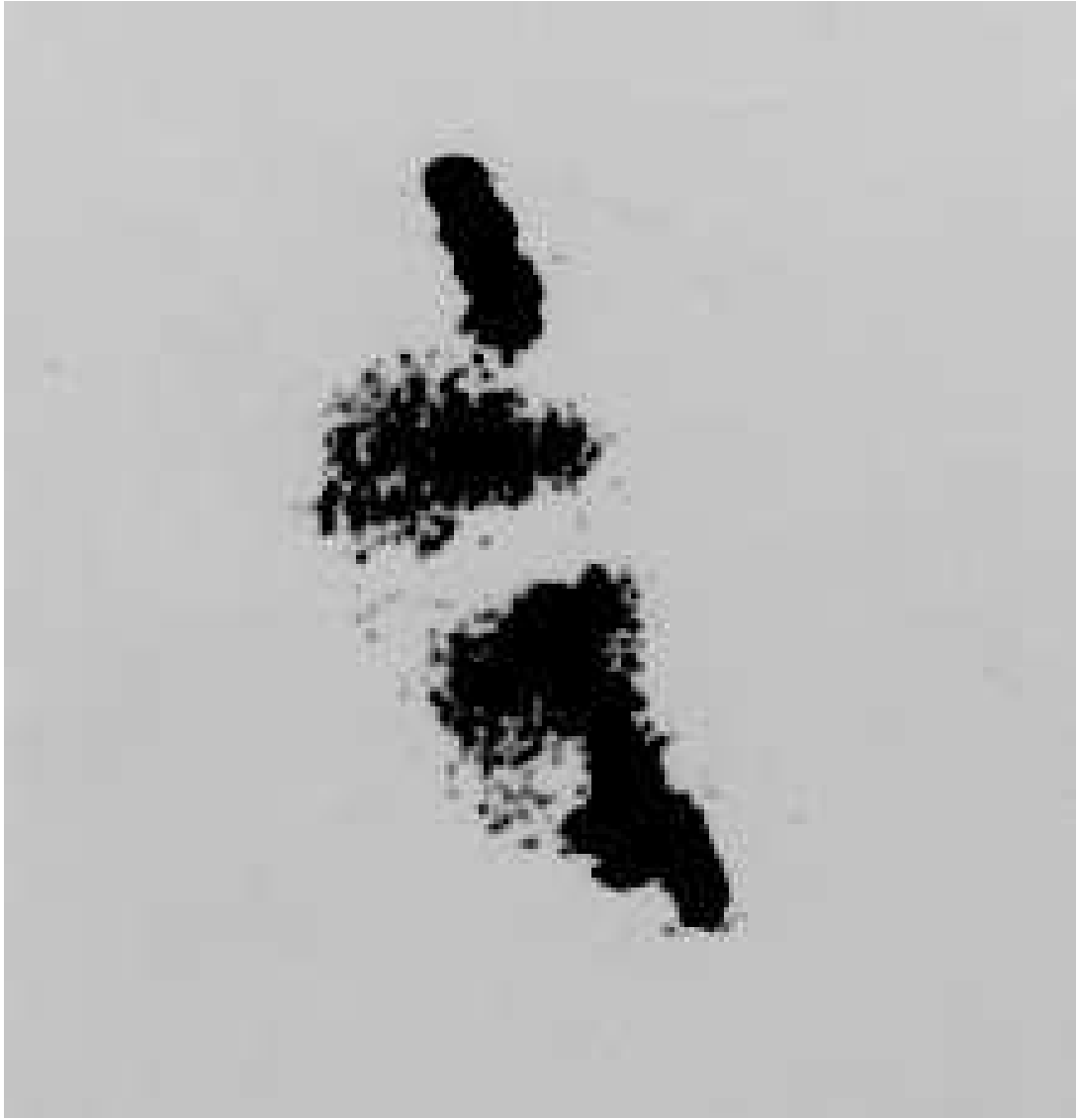 ,height=15.00cm}}
\caption{\fc Total intensity radiograph for 3C33 at 2 cm, made with the VLA,
showing the sharp inner boundaries of the lobes.  
Reprinted with permission from Leahy \& Perley, AJ, 102, 537 (1991). \copyright 
American Astronomical Society}
\end{figure}
The sharp-edged central emission gaps can be attributed to the blocking 
of `back-flowing' nonthermal plasma of the radio lobes
by a denser thermal plasma associated with the parent galaxy. Thus, from
the `height' of the radio gaps in these 12 sources ($z_{\rm median} \sim 0.2$) 
we estimated a median value of $\approx 75~$kpc for the disk diameter, 
which is clearly a lower limit. Likewise, the observed `widths' of the gaps suggest
a median value for the disk thickness, $l_{disk} \approx 25~$kpc. Such enormous
disks ({\it superdisks}) are consistent with the numerical simulations of 
the galaxy formation at high redshifts in hierarchical clustering scenarios
(e.g., Evrard, Summers \& Davis 1994). Since 
quasars at $z > 1$ have a median radio size of $\sim$60--70 kpc (taking
$H_0$ = 75 km s$^{-1}$Mpc$^{-1}$ and $q_0~=~0.5$), {\it their far-side lobes could
well be largely hidden behind the postulated superdisks}. 

Gopal-Krishna \& Nath (1997) have discussed a scenario for the thermal
evolution of the tidally-stretched gaseous disk of the captured galaxy, as
it gravitates further toward the interior of the hot gaseous corona
of the massive elliptical. They estimate that for plausible gas densities, the
disk would continue to tap heat from the ambient coronal gas and expand in
width to $\approx 10 $kpc in $10^8$--$10^9$yr, which is the expected formation
timescale for the dust-lanes. Such widths are consistent with the observed 
dimensions of the gaps along the radio bridges. Interestingly, direct evidence 
for depletion of cool interstellar medium (HI) of a captured disk galaxy, {\it
 via} heat input from the coronal gas of the captor galaxy, has come from  
 HI imaging of the early-type galaxy NGC1291 (Bregman, Hogg \& Roberts 1995). 
Note also that in addition to widening it, the heat input into the disk is also expected 
to steadily increase the diameter of its ionized central part, 
concurrently with the growing size of the radio source. 
The estimated average disk parameters 
(magnetic field, $B \sim~1 {\mu}$G, and electron column density, $\Sigma~
\simeq~300~$pc cm$^{-3}$) correspond to a Faraday dispersion parameter,
$\Delta \approx \langle n^2 B^2 \rangle^{1/2}$ $(l_{disk})^{1/2}$ $\simeq~10^2$cm$^{-3} \mu$G pc 
(Gopal-Krishna \& Nath 1997; see also Gopal-Krishna \& Wiita 2000 -- hereafter GW00). 
Such values are quite consistent 
with those inferred from the depolarization maps of quasars (Garrington \& 
Conway 1991). 
Sensitive UV/X-ray
imaging with space observatories could possibly provide strong support of this
scenario, via a direct detection of the hot component of the postulated 
magneto-ionic superdisks
around powerful, distant radio galaxies, although the combination
of spatial resolution and sensitivity seems to be just beyond the
current generation of X-ray telescopes ({\sl Chandra} and {\sl XMM--Newton}). 
Perhaps a more promising strategy
to detect such disks would be to map the dust-reddening of the background
galaxies.
\\

\noindent{\s 4. The radio spectral-index asymmetry}
\\

\noindent
While the depolarization asymmetry (Laing-Garrington effect) can be 
naturally explained in terms of orientation effects (\S 3),
there is still no consensus on the origin of another prominent asymmetry 
which involves the radio spectral index. This intriguing (statistical) asymmetry 
was noticed in two separate studies of 3CR radio sources, and can be 
summed up as follows:

\noindent (a) The lobe which is less depolarized has a flatter radio spectrum (Liu
\& Pooley 1991b).

\noindent (b) The lobe on the jet-side has a systematically flatter spectral index
    by typically $\Delta\alpha~\sim~0.1$, though with a large dispersion (Garrington et al.\ 1991).

Note that the first statement refers to a sample consisting chiefly of radio
galaxies, for which orientation effects are expected to be weak, though
not necessarily negligible.
Specifically, for such sources the longer lobe tends to be less 
depolarized, which is consistent with the halo of magneto-ionic plasma 
having a decreasing radial profile (e.g., Pedelty et al.\ 1989; 
also, Strom \& J\"agers 1988).

As argued by Liu \& Pooley (1991a,b), the effect (a) does not arise from the jet 
contaminating the approaching lobe's emission, since the jet contribution would
be inadequate. These authors also discount the possibility that pure Doppler
shifts, in conjunction with a downward curved spectrum of the radio lobes could
explain the observed spectral asymmetry (see, also, Blundell \& Alexander 1994). 
Nonetheless, Doppler effects could lead to strong flux boosting of 
the approaching hot-spot, thus raising its contribution to the flux density 
of the nearer lobe (which is associated with the observed jet). This is also 
consistent 
with the result that the hot-spot on the jet side is usually brighter (Bridle
\& Perley 1984; Laing 1988; Teerikorpi 1984; Dennett-Thorpe et al.\ 1997). Since the hot-spot spectrum is flatter than
the lobe spectrum, the enhanced hot-spot contribution would flatten the
integrated spectrum of the nearer lobe (which is associated with the 
brighter jet and
is usually less depolarized), as compared to the farther lobe (Garrington et
al.\ 1991). Extending this argument, Tribble (1992) estimated that the approaching 
hot-spot {\it itself} might exhibit a measurably flatter
spectrum compared to the receding hot-spot, since the core of the approaching
hot-spot can be expected to have not only a flatter spectrum than the 
annular portion, but also a greater Doppler boost due to a faster
forward motion (which, incidentally, could be well above the time-averaged
speed of the hot-spot; Scheuer 1982). One may thus be able to understand 
not just the compactness asymmetry of the hot-spots (Laing 1995; see Tribble 
1992) but also their spectral-index asymmetry, as reported by Liu \& 
Pooley (1991a,b). Tribble has further argued that 
the rather strong correlation between the depolarization asymmetry
and the spectral-index asymmetry is an indication that the latter, too, is 
largely an orientation effect.
\\

\noindent{\s 4.1. A further twist}
\\

\noindent
However, another study calls for caution in accepting the
apparently attractive explanation of the spectral-index asymmetry of the lobes 
in terms of a greater contamination from the brighter,
Doppler-boosted hot-spot present on the jet side (see above). From detailed
VLA observations of a set of 8 powerful radio quasars, Bridle et al.\ (1994)
have shown that nearly in every case,  the
hot-spot  on the jet side has {\it itself} a flatter spectrum than the hot-spot 
on the counter-jet side. Moreover, the hot-spots on the jets' sides do not 
appear systematically brighter. While it is important to settle these points
conclusively by mapping a fairly large sample of quasars with the VLBA, a few 
interesting possibilities emerge from the work of Bridle et al. 
Basically, they find that whereas the radio spectrum of high surface 
brightness features (hot-spots) is almost always flatter on the jet-side,
the diffuse lobe emission on that side has a {\it steeper} spectrum, if the
associated lobe is significantly shorter than the other lobe. 
If true,
this could imply a link between the mechanisms of producing a shorter lobe on 
the jet side (which is unexpected in the simple kinematical model; \S 2.1)
and the energy loss rate of the radiating electrons within the lobes. 
It is clearly important to establish the generality of the result of Bridle 
et al.\ (1994), which suggests
that while the jet-side controls the spectral asymmetry of the bright compact
emission, the spectral asymmetry of the diffuse emission is controlled by
the lobe-length asymmetry. 

An `intrinsic' explanation for the spectral asymmetry of the hot-spots/lobes
on the two sides has been proposed in terms of a steady dilution of the
magnetic field as the radio source grows, essentially in a self-similar 
fashion (Blundell \& Alexander 1994). From a quantitative, 
albeit highly model-dependent, treatment based on this idea, these authors 
are able to explain the typically observed spectral asymmetry  
($0 < \Delta\alpha < 0.4$). They argue that, due to projection and 
differential light travel time effects, 
the lobe on the counter-jet side is monitored at an earlier stage, when its
magnetic field is postulated to have been stronger and hence its spectrum
steeper, thanks to heavier radiative losses. This simple scenario predicts: 
(i) 
a higher surface brightness for the lobe with a steeper spectrum and stronger
depolarization;  and (ii) the magnitude of the spectral asymmetry should be 
maximum for the outer edges of the lobes and minimum for the parts close 
to the nucleus. Subsequent observations 
by Dennett-Thorpe et al.\ (1997) did not bear out this prediction, as they
found that for more aligned jets, the near hot-spot has a higher flux
and a flatter spectrum; however, their 
sample is small (just 6 sources with detectable jets). Interestingly, these
authors find that in 5 of the 6 sources the lobe with a larger surface area
has a flatter spectrum. Thus, at this stage, it remains a challenge to disentangle the 
intrinsic, environmental and orientational effects in the interpretation of 
the spectral asymmetry, and only larger samples will allow these
important tasks to be accomplished.
\\

\noindent{\s 4.2. Correlation of the structural asymmetry with spectral curvature }
\\

\noindent
Yet another  type of asymmetry was noticed from a careful analysis
of the radio spectra of a well-defined large sample of FR II radio galaxies
selected from 3CR sources. The available data for these galaxies allowed 
the determination of the {\it rest-frame radio spectrum} for individual sources,
after subtracting the compact radio emission associated with the core.
It was thus found that, statistically, the lobe spectra are more curved for 
sources with larger core-to-hot spot separation ratio ($Q$) (Mangalam \& 
Gopal-Krishna 1995). This correlation between 
spectral-curvature ($C$) and asymmetry ($Q$) needs to be further investigated using 
a larger sample, as it can provide potentially useful clues about the 
evolutionary tracks for powerful radio sources. 

It is instructive to consider the $Q$--$C$ correlation in conjunction with 
the result that a larger spectral curvature is more prevalent among 
FR II sources having normal spectra (i.e., not very steep) near the 
rest-frame frequency of 1 GHz (Mangalam \& Gopal-Krishna 1995). 
This is suggestive of a temporal evolution heading toward not only
a steepened but also a straightened 
spectrum (within the observed radio window), saturating
near $\alpha \approx -1.2$ (Mangalam \& Gopal-Krishna 1995). Since such 
an evolution of powerful radio sources could arise from an accumulation of 
energy losses, the source asymmetry should be linked to this factor, 
in view of the $Q$--$C$ correlation. One possibility is that within a
radius of $\approx10~$kpc of the nucleus the circum-galactic medium is not
only dense but also very clumpy. At least for distant RGs at
$z~\ge~2$, an abundance of cool gaseous clumps in the circum-galactic medium
has been inferred from the observed multiple absorption troughs in the extended 
Lyman-$\alpha$ emission profiles of RGs with dimensions
up to $\sim50~$kpc (van Ojik et al.\ 1996).  
The presence of such a clumpy circum-galactic medium can have a profound
influence on the early evolution of radio sources, leading to a conspicuous 
structural asymmetry and, concurrently, a curved radio spectrum, by slowing 
down the jets and thereby prolonging the efficient confinement of the lobes 
(Mangalam \& Gopal-Krishna 1995; Saikia et al.\ 1995; Wang, Wiita \& Hooda
2000; Mellema, Kurk \& R{\"o}ttgering 2002).

An interesting possible link between the structural asymmetry and the 
presence of a dense ISM in the parent galaxy is hinted by the work of Cimatti
et al.\ (1993).  They found a tendency of markedly asymmetric radio
sources to show substantially polarized light across the parent galaxy; the 
polarization probably results from scattering of the nuclear light by a
dense ISM of the galaxy.  
\\

\noindent{\s 5. Correlated radio-optical asymmetries in powerful radio galaxies}
\\


\noindent
A  major impetus for linking the environmental factor to the asymmetries has 
come from the study of the extended optical emission-line regions (EELR)
associated with the FR II type 3CR radio galaxies, as reported by McCarthy \& 
van Breugel (1989); MvBK (see, also, Liu \& Pooley 1991b).
 These authors found a {\it very strong} tendency for the optical
surface brightness to peak on the side of the shorter radio lobe. Further, the
optical brightness contrast between the two sides was found to be rather
pronounced for higher redshift sources ($z~>~0.3$) for which the line emission 
was measured in terms of the blue line [O II]$\lambda$3727. MvBK proposed 
that the causal link between these tightly correlated radio and optical 
asymmetries is 
a large-scale density asymmetry in the circum-galactic medium of the
radio-loud elliptical, which slows the lobe propagation more effectively
on the denser side (e.g., Hintzen \& Scott 1978; Swarup \& Banhatti 1982),
while simultaneously enhancing the optical output of the thermal filaments
of the EELR on that same side {\it via} a stronger confinement of the filaments. 

Although appealing for its simplicity, this hypothesis invoking environmental 
asymmetry does not seem to offer a robust explanation for the {\it nearly
universal} correlation which is found to hold even when the lobe-length 
asymmetry is at a level of just $\approx 1\%$ (MvBK),  particularly in view of the
fact that in a given lobe the physical separations of the radio and optical 
peaks from the nucleus usually differ by an order-of-magnitude (MvBK). 
On the other hand,
should the spatial scale of the environmental density asymmetry indeed be large 
enough to encompass the entire radio lobe, the putative excess ambient 
pressure enhancing the optical line-emission from the shorter lobe through a
more effective confinement on that side should also enhance that lobe's radio
output. No significant trend of this sort is noticed, however
(MvBK). Furthermore, it remains unclear why the optical
asymmetry is more pronounced for higher redshift galaxies whose EELRs were
imaged in the blue [O II]$\lambda$3727 line, as compared to the low-redshift 
galaxies imaged in the (red) H$\alpha$ line (MvBK). Note also that if the 
trajectories of the twin jets are not precisely anti-parallel, or linear
(which is likely, given the ``dentist's drill" behavior of the hot-spots;
e.g., Scheuer 1982; Clarke 1996; Norman 1996), then the asymmetry parameter, $Q$, could 
easily depart mildly from unity, without there being any systematic 
difference present between the two lobes' environments. These various 
difficulties have
led us to conclude that invoking a nearly universal environmental density 
asymmetry about radio-loud ellipticals, in order to explain the tight 
radio-optical correlated asymmetry is probably 
too simplistic, and a {\it more foolproof} mechanism is likely to be at work 
(Gopal-Krishna \& Wiita 1996; hereafter GW96).
\\

\noindent{\s 5.1. A new scenario for the correlated radio-optical asymmetry and
the lobe-length asymmetry}
\\


\noindent
Our proposal is based on the premise that the motion of the parent galaxy 
against the external medium will cause a ram pressure sweeping of the
(dusty) ISM, leading to its asymmetric disposition about the galaxy.
Consequently, the dust obscuration of the nuclear UV photons
and of the light emitted from the EELR filaments will become stronger and
more widespread on the downstream side of the galactic center. At the same
time, the radio lobe on that side would be automatically identified as the
longer one, due to the continuing motion of the galaxy in the opposite
direction. Not only would this naturally engender the correlated radio-optical
asymmetry, but it would also explain the strong tendency for the EELR asymmetry to 
be more marked at higher redshifts, since the observed optical window would
then sample shorter rest-frame wavelengths, where the dust obscuration is
expected to be more pronounced. Details of this scheme are spelled out in 
GW96 (see, also, \S 5.3).

Using a sample of 12 radio galaxies showing evidence for a 
superdisk, GW00 noted an interesting trend: the 
lobe-length asymmetry is systematically
milder when the lobe lengths are measured from the mid-plane
of the radio emission gap (the superdisk), instead of 
employing the usual definition when the lobe lengths are
measured using the radio nucleus
as the reference point. The
offset of the radio nucleus from the superdisk midplane thus
indicates a motion of the host galaxy relative to the superdisk,
which is possibly a portion of a cosmic filament.
Thus, within a typical age of a radio source, $\tau~> 10^{8}$yr
(\S 5.2),
the parent galaxy would have moved a distance $D~=~30\tau_8 V_{g(300)}$kpc from
the origin, marked by the point of symmetry between the two hot-spots, 
where the lifetime $\tau_8$ is in units of $10^8$ years and
 the galactic velocity component along the radio axis,
$V_{g(300)}$, is in units of 300 km s$^{-1}$. Based on the results of various
published simulations, GW96 adopted $V_{g(300)} \sim 1$ for the powerful radio 
sources at $z > 0.5$, which tend to occur inside clusters of galaxies (Yee \&
Ellingson 1993, and references therein). Now, a $D$ of $\sim 30~$kpc
corresponds to an asymmetry parameter $Q \sim 2$, given that the typical
size of powerful RGs at $z > 0.5$ is $\sim$150 kpc. 
Since, for distant RGs the observed $Q_{\rm median} \sim~1.5$,
it seems that the motion
of the parent galaxy should play a very significant role in causing the 
lobe-length asymmetry, though some rare, highly asymmetric, sources 
would still warrant  alternative explanation(s).  
\\

\noindent{\s 5.2. Estimated lifetime of nuclear activity in double radio sources}
\\

\noindent
Historically, the ages of the radio sources have been determined from 
the observed large-scale spectral gradients along the lobe's length, by 
taking into account the main energy loss processes, involving radiative 
and expansion losses of the energetic particles.  The ages thus derived 
are typically of order of $10^7$ years (e.g., Leahy, Muxlow \& Stephens 
1989; Liu et al. 1992). However, a few authors have 
questioned the various assumptions underlying this method (e.g., Rudnick 
et al.\ 1994; Blundell \& Rawlings 2000, 2001;
see Wiita \& Gopal-Krishna 1990,
for one of the earliest criticisms of the spectral ageing method). 
For instance, in their critique, Blundell  \& Rawlings  (2001) have argued that the
ages may have been grossly underestimated on account of the following 
(unrealistic) assumptions:
1)  temporal constancy of magnetic field in the lobes;
2)  power-law nature of the electron injection spectrum into the lobes;
3)  radiative lifetimes of the synchrotron emitting particles in the
lobes are longer than the ages to be deduced;
4)  no mixing of the relativistic particle population within the lobes.

By modelling a several flux-limited samples, totalling 303 radio sources 
selected near 
0.15 GHz, their best-fit to the data in the 
spectral index--size--luminosity--redshift ($\alpha-P-D-z$) space yielded 
a typical activity lifetime of $5 \times 10^8$ years (Blundell, Rawlings
\& Willott 1999). These longer 
lifetimes are in better accord with the ages inferred from the hot spot 
velocities estimated from lobe length ratios (e.g., Scheuer 1995; \S 2.4). 
Because of severe inverse Compton losses to the microwave
background at high redshifts, very few large, therefore old, sources
are  detectable at higher $z$ (Gopal-Krishna et al.\ 1989); this was an
early version of the so-called ``youth-redshift degeneracy'' (Blundell et al.\ 1999).

More recently, independent support for such long nuclear activity has come 
from several studies made in other wavebands, and we now note a few of
these investigations.  Employing very deep VLA observations of a hard X-ray selected sample of
   69 sources, Barger et al.\ (2001) have found that almost
   4\% of the $L > L_\ast$ galaxies are X-ray luminous at any 
   time, indicating an average activity lifetime of $\sim 5 \times 10^{8}$ 
years for their supermassive black holes.
Using a large sample of $\sim 5000$ massive galaxies at $z < 0.1$,
   detected in the SDSS Survey (York et al. 2000), Miller et al.\ (2003)
   have found that almost 40\% of them contain an AGN. From the high
   fraction of AGN among the massive black holes in the local universe,
   they infer an average activity lifetime to be $> 10^8$ yr.
Most recently, Marconi et al.\ (2004) have estimated the local ($z$ = 0)
   distribution of black hole mass to be $\sim 5 \times 10^{5} M_{\odot}{\rm Mpc}^{-3}$.
   Interpreting these as exclusively the relics of AGN activity (since 
   $z < 3$), they estimate an average total activity lifetime to be 
   in the range $10^8 - 10^9$ years.

These new data supporting long periods of RG activity add
additional weight to our argument that galaxy motion is likely
to produce a substantial portion of the observed asymmetries (\S 5.1).
\\
\newpage
\noindent{\s 5.3. Asymmetric {\it obscuration}, instead of asymmetric {\it emission}}
\\
 
\noindent
While careful optical/UV observations have strengthened the case for
distributed dust within the ISM of RGs and quasars (e.g., Goudfrooij
\& de Jong 1995; Wills et al.\ 1993),  evidence is also emerging for extended dusty 
superdisks in such objects, as summarized in \S 3.1. 
The latter can act as a rich reservoir of dust which is gradually released 
into the diffuse ISM on the downstream side, as the galaxy moves on. 
GW96 have discussed evidence
for the galaxy motion, based on the observed offsets of the radio nucleus
from the mid-point of the radio emission gap perceived to coincide
with the dusty disk. As argued in GW96 and GW00, the 
combined effect of the dust sweeping from the ISM and the disk component into 
the downstream side of the moving host galaxy would be to enhance the degree and
domain of obscuration on that side {\it vis-\`a-vis } the upstream side which 
would {\it necessarily} coincide with the shorter radio lobe. The result 
would be a
diminishing of the magneto-ionic content, and especially a decline in the EELR 
emission, on 
the side of the longer lobe. This is expected, in that an enhanced 
dust attenuation of the nuclear ionizing photon flux would be compounded by the 
attenuation of the UV/blue light emerging from the EELR. 

To summarize, in our picture, the tightly correlated radio-optical asymmetry 
of powerful RGs is not contingent upon the conventional assumption 
of a nearly universal density asymmetry about the host galaxy. Instead, the
correlated asymmetry is driven largely by the galaxy motion which leads 
to asymmetric obscuration on the
two sides. The various difficulties with the conventional explanation can thus 
be circumvented (see above). These include: (i) the radio-optical asymmetry 
correlation being present even if the lobe-length ratio departs only minutely from 
unity, in spite of the radio and optical emission originating on 
vastly different physical scales;
(ii) the asymmetry being much more pronounced in the (blue)
line [O II]$\lambda$3727, as compared to the (red) H$\alpha$ line;
and (iii) the tendency of the shorter lobe to be frequently the fainter one in the
radio, though the opposite would always be expected in case the lobe-length 
asymmetry were firmly rooted in an environmental asymmetry.  Variants of this
idea and additional implications are discussed in GW00.
\\

\noindent{\s 6. Lyman-$\alpha$ halos as the diagnostic of dusty {\it superdisks}
in high-$z$ RGs}
\\


\noindent
A prime motivation to search for galaxies at high redshifts has been the 
prospect of detecting their extended emission components which, in the case of 
(distant) quasars, are usually swamped by the bright (point-like) nuclear 
emission.  Furthermore, the giant Ly-$\alpha$ nebulosities found associated 
with many distant RGs can provide a sensitive probe of dust 
in the early universe, since Ly-$\alpha$ photons, being resonantly trapped 
within the nebula, can be efficiently  absorbed by even a small amount of dust. 
Moreover, as discussed by Gopal-Krishna et al.\ (1995) and GW00, any conspicuous
brightness 
asymmetry of the Ly-$\alpha$ nebulosity about the galactic nucleus could 
often be useful for distinguishing the near-side lobe from its counterpart
on the far-side of the nucleus; making this distinction normally remains 
only a cherished dream, 
since radio jets  often remain undetected in powerful RGs.

\begin{figure}
\centerline{\psfig{figure=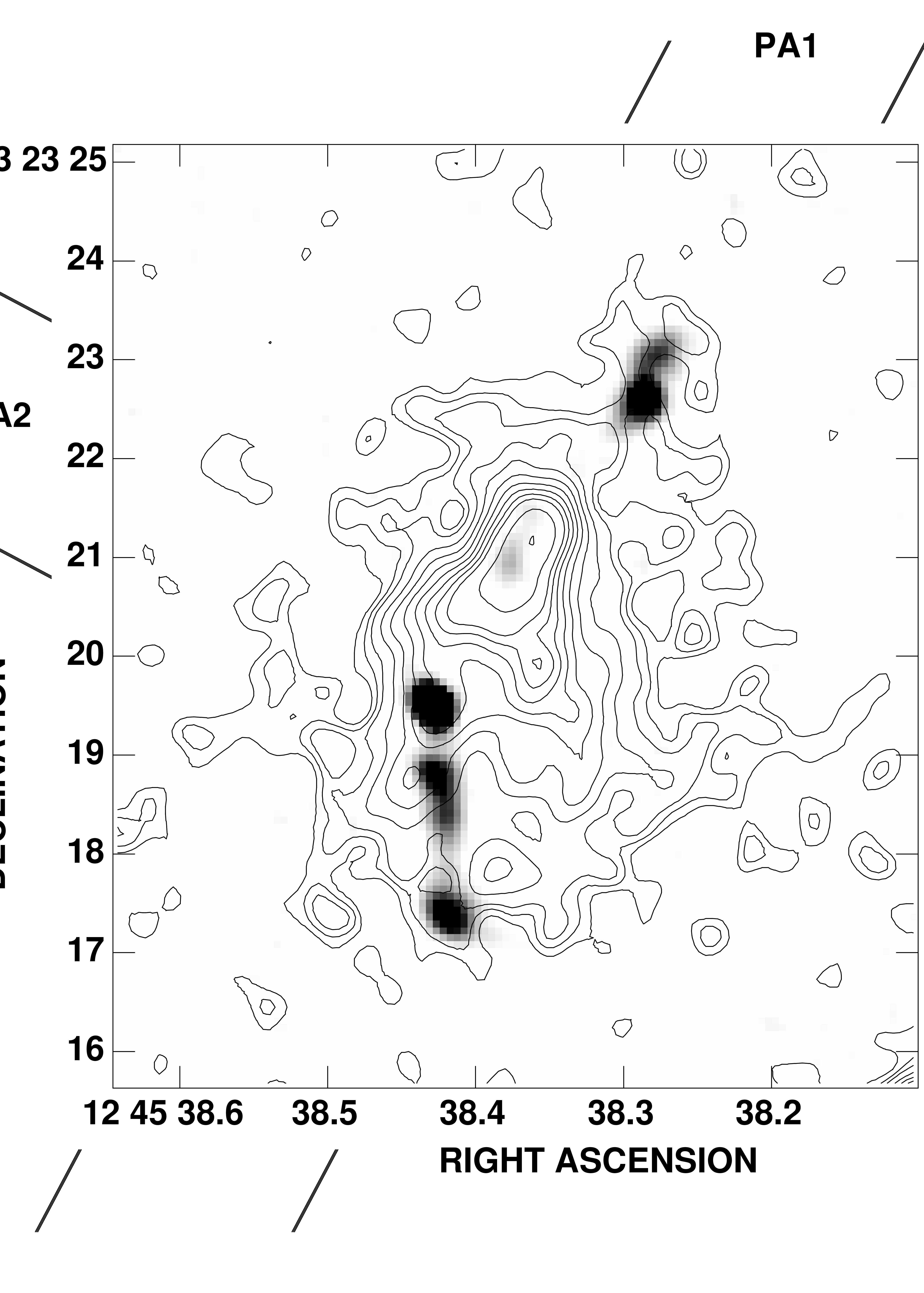 ,height=15.00cm}}
\caption{\fc The Ly$\alpha$ image of 1243+036 in contours
with the 8.3 GHz VLA map in greyscale superimposed. The bending point of the 
radio structure coincides with enhanced
Ly$\alpha$ emission and a steep rise in the line-width in one of the `arms'. 
Indicated around the edges
are the slit positions that were used in the high
resolution spectroscopy. Reproduced with permission from van Ojik et al.\ 
(1996). \copyright European Southern Observatory}
\end{figure}

Fig.\ 2 shows a  giant Ly-$\alpha$ cloud, which is associated
with the radio galaxy 1243+036 at $z~=~3.6 $ (van Ojik et al.\ 1996).
Several more examples are discussed by Gopal-Krishna et al.\ (1995)
and GW00.
A plausible explanation for most such clouds is that the ejection 
of radio jets, probably triggered by the accretion of dusty material onto
a collapsed object (e.g., Lynden-Bell 1996), blows an expanding radio cavity
through a pre-existing giant circum-galactic nebula with a two-phase
temperature/density structure (Rees 1988). This creates a cocoon
of nebular gas around the cavity, which is compressed, heated and accelerated from 
within by the expanding radio lobes and their bow-shocks. Denser clumps
in this ionized cocoon manifest themselves as the Ly-$\alpha$ nebula. (A similar
model was proposed by Meisenheimer \& Hippelein (1992) to explain
the [O II]$\lambda3727$ nebulosity around the radio galaxy 3C368.) The
importance of radio lobes in the creation of giant Ly-$\alpha$ nebulae 
is further
underscored by the finding of Heckman et al.\ (1991b) who noticed a
lack of resolved Ly-$\alpha$ emission among quasars with under-developed 
radio lobes (i.e., $\le 10$kpc). More recently, Giraud et al.\ (1996) 
have argued that nuclear UV photons can be an important, and probably dominant, 
source of excitation for even such giant nebulae (see, also, van Ojik 
et al.\ 1996).  

In the now standard hierarchical models of the growth of cosmic structure,
such Ly-$\alpha$ clouds and newly forming galaxies will be portions
of the web of sheets and filaments of baryonic matter collecting within
dark matter halos (e.g., Dav{\'e} et al.\ 2001).
\\

\noindent{\s 6.1. Asymmetry of Lyman-$\alpha$ halos as a clue to the sidedness}
\\

\noindent
Discounting the virtual impossibility that in all radio galaxies, the disk
is perfectly planar and, moreover, viewed precisely edge-on, one expects 
to find
cases where the Ly-$\alpha$ halo associated with the 
far-side radio lobe is at least partly hidden behind the disk, and is thus 
effectively obscured.  If, indeed, the Ly-$\alpha$ brightness asymmetry
about the nucleus is because the brighter part of the cloud is associated
with the {\it approaching} radio lobe (and hence not obscured by the central
dusty disk), then one would expect to detect in the kinematics of the 
Ly-$\alpha$ cloud a signature of the lobe's motion in our direction. 

Such a distinct signature, namely, a negative velocity gradient (blueshift)
along the Ly$-\alpha$ cloud from the nucleus toward the outer edge of the 
lobe, has indeed been found in the case of those high-$z$ RGs
which exhibit a distinctly asymmetric Ly-$\alpha$ brightness about the
nucleus {\it and} for which a velocity profile along the cloud is available.
 These galaxies are: 3C294 ($z$ = 1.82, McCarthy et al.\ 1990), OTL 0852+124 
($z$ = 2.47, Gopal-Krishna et al.\ 1995), 8C 1435+635 ($z$ = 4.25, Lacy et al.\
 1994; Spinrad, Dey \& Graham 1995) and 1243+036 ($z$ = 3.57, 
van Ojik et al.\ 1996). The case of the $z$ = 3.8 galaxy 4C41.17 is somewhat 
confusing, however. The brightest part of the Ly-$\alpha$ cloud which
extends to the northeast of the nucleus by $\sim 3^{''}$, appears, instead,
to be {\it redshifted} relative to the nucleus, though the velocity gradient 
reverses still further to the east (Hippelein \& Meisenheimer 1993; Chambers, 
Miley \& van Breugel  1990). Nonetheless, the detection of a radio jet on
the eastern side of the nucleus (Carilli, Owen \& Harris 1994) strongly 
supports our case that the brighter Ly$-\alpha$ halo seen on the eastern 
side is associated with the approaching radio lobe. Further discussion on
this point can be found in GW00.
\\

\noindent{\s 7. Powerful radio sources with one-sided radio structure?}
\\

\noindent
If indeed, going by the standard wisdom, radio lobes have no bulk 
relativistic motion, the existence of sources with their entire radio 
structure on just one side of the parent galaxy would furnish strong 
evidence for a central engine capable
of ejecting a one-sided jet (\S 7.2). As a class, such one-sided radio quasars 
were first highlighted and christened `D2 quasars' by Macdonald \& Miley (1971). 
In subsequent deeper radio imaging of the candidates, the `missing' lobe has
been spotted in most cases (e.g., Saikia et al.\ 1990, and references
therein), strengthening the possibility that the D2 quasars  simply populate 
the tail of the lobe flux-ratio distribution. That the extreme flux ratios 
may actually arise from orientation dependent factors is intimated by the 
scarcity of such sources among RGs and lobe-dominated radio sources,
in general
(e.g., LHG; Saikia et al.\ 1990; Browne \& Perley \ 1986). 
\\

\noindent{\s 7.1. The problem of 3C273}
\\

\noindent 
One source which continues to queer the pitch by steadfastly frustrating all 
attempts to detect any lobe 
associated with the putative counter-jet is 3C 273, 
the prototypical D2 quasar.  In spite of being imaged with a dynamic range of 
up to 10,000:1, no convincing detection of a counterpart to the main radio 
lobe has been possible (Conway et al.\ 1981; Davis et al.\ 1985). Is it then 
the rare example of 
truly one-sided jet? Or, alternatively, is in this source even the lobe 
advancing relativistically and at a small angle to the line-of-sight? Since 
this may seem unlikely, in view of the rather inconspicuous appearance of
the lobe even around the approaching jet, a further alternative has been 
suggested which obviates the need for a highly relativistic lobe 
velocity, while still admitting the usual two-sided jet ejection scenario. 
Basically, in this scenario, 3C 273 is envisioned to be a very energetic, 
young ($\sim10^{6}$yr) source, with its highly collimated beams,
pointed close to the line-of-sight ($\theta \le 20^o$) and advancing rapidly
($\beta \ge 0.6)$. Such a rapid advance of the beam is not 
conducive to lobe formation. Furthermore, due to the transit-time delay, any
lobe associated with the tip of the putative counter-jet would be monitored
when it was a few times younger than the observed approaching lobe, i.e.,
at a stage which was probably too early for lobe formation 
(Kundt \& Gopal-Krishna 
1981, 1985; Morrison, Roberts \& Sadun 1984; Morrison \& Sadun 1992; Stawarz
2004).  Since this explanation requires 
several special circumstances, including a close orientation, young age, and a
fairly rapid jet advance, the extreme rarity of 3C273-like objects
 would not be too surprising.
\\

\noindent{\s 7.2. Producing intrinsically one-sided jets}
\\

\noindent
Although only about one-third of all well mapped radio sources exhibit 
both jets on the kiloparsec-scale, while another third allow us to view 
one jet, and the remainder, neither (e.g. Bridle \& Perley 1984),
the idea that all sources contain essentially symmetrical pairs of jets
is inspired by the global symmetry of their radio lobes. The twin-beam picture
of Blandford \& Rees (1974) suggested that an initially isotropic ejection
of hot relativistic plasma from a galactic core
could be channelled through de Leval nozzles established in a 
flattened surrounding
medium, into a pair of equally powerful jets.  Though the discovery from VLBI
of one-sided jets, already on the parsec-scale, showed that
this picture was probably not relevant on the $\sim 100~$pc scale, as
envisioned by Blandford \& Rees, the possibility that this mechanism
played some role on smaller scales could not be excluded (Wiita 1978).
Even if the jets are produced in the immediate environs of the assumed
supermassive black holes and are intrinsically symmetric on that (innermost)
sub-pc scale, interactions with a possibly asymmetric ambient
medium on a wide range of scales could easily modify the effective
thrusts and/or opening angles of the jets, thus engendering an 
intrinsic asymmetry.

Wiita \& Siah (1981) demonstrated that even a modest offset of the
central source from the center of mass of a confining ISM cloud might choke
or greatly weaken one of the jets, while allowing the other to burst out.  The
possible orbiting of the jet emitting engine about the center of the confining
cloud could imply that the jets emerged alternately on one side and then 
on the other, yielding the rough symmetry as a time-averaged effect 
(e.g., Saikia \& Wiita 1982). Icke (1983)
suggested a hydrodynamical ``clamshell'' mechanism that could possibly
produce such inversions in the central regions of galaxies.
Rudnick \& Edgar (1984) reported a significant number of sources with
asymmetric radio features that appeared to mirror this ``flip-flop''
phenomenon (see, also, LHG). The possibility of a binary black hole
system in the active nucleus has also been proposed as a way of producing
alternating side ejections (Begelman, Blandford \& Rees 1980;  Komberg 1994).
The apparent need for continuous reacceleration of
synchrotron emitting electrons because of their short life-times is, however, 
a strong argument in favor of the
vast majority of sources being continually fed on both sides, even if
not with a perfect balance.
 
A discussion of the various mechanisms proposed for the launching and
initial collimation of the jets is beyond the scope of this article
(see Wiita 1990, 1996 for reviews); here we merely note that
essentially all those models are based on accretion disks around
supermassive black holes, invoke magnetic fields and tend
to produce quite symmetrical, oppositely directed jets.
Still, the possibility that asymmetric jets could be produced by
magnetically dominated mechanisms has been discussed (e.g., Bodo et
al.\ 1990; Bisnovatyi-Kogan \& Moisenko 1992).  The most detailed
proposal along these lines 
is the model of Wang, Sulkanen, \& Lovelace (1992), which allows for
magnetized disks asymmetric about the mid-plane that can produce
very asymmetrical jets.  The Wang et al.\ models are
field-dominated and carry energy away from the disk mostly in the
Poynting flux of electromagnetic fields.

Other evidence for intrinsically asymmetrical jets is presented by
Fraix-Burnet (1992), who points out that Doppler favoritism may not be
adequate to explain one-sided jets in sources like M87, NGC 6251 and
Cygnus A, whose jets are probably oriented close to the plane of the sky
(see, however, Jones 2000).
Fraix-Burnet (1992)  argues that while a pair of jets is probably
always launched, in many cases only one is observed. In stark contrast
to the usual assumptions about the consequences of jet interactions with
external media (\S \S 5, 7.5), in this scheme the visible jet is
supposed to be interacting less with the ISM than is its invisible counterpart. 
The idea behind this radical interpretation is that if magnetohydrodynamical
turbulence produces most of the synchrotron emission from a jet, it may
be enhanced if the boundary layer is thinner, i.e., when the jet does
not interact intensively with the ISM.
\\

\noindent{\s 7.3. The apparent asymmetry of jets in powerful radio sources}
\\

\noindent
The asymmetry of optical jets on the kiloparsec-scale was the first striking 
illustration of asymmetry in RGs (i.e., M87), even preceding
the advent of radio astronomy (\S 1). The initial explanation invoking a bulk 
relativistic motion of the radiating plasma has received mounting 
support from observations in different wavebands (Begelman, Blandford \& Rees
1984; Scheuer 1987; Urry \& Padovani 1995; Cohen \& Kellermann 1995;
Kellermann et al. 2004). Some of the well known evidences for relativistic 
motion in both parsec and kiloparsec-scale jets include:  
the apparent superluminal motion of the parsec-scale jets and their 
correlated sidedness with the {\it kiloparsec}-scale jets; the
apparent deficit of inverse Compton boosted X-rays from the compact jets;
the lobe-depolarization asymmetry of quasars (\S 3); the intra-day radio
variations; the rapid time variability of intense $\gamma$-ray emission 
from blazars; and the correlation of radio structural asymmetry and distortion 
with the core prominence.  Instead of repeating those extensive arguments here, 
we refer the readers to some review article (e.g.\ Bridle 1996; Laing 1995; 
Bicknell 1995; Urry \& Padovani 1995; Gopal-Krishna 1995; Chiaberge 2004). 
Additionally, we recall the growing recognition of the free-free absorption 
by the nuclear disk as a possible cause of the apparent asymmetry of
the parsec-scale radio jets (e.g., Vermeulen, Readhead \& Backer 1994; Romney 
et al.\ 1995; Matveyenko et al.\ 1980;\ 1996). 
  
Here, we briefly recount some observational clues which serve to reveal 
(indirectly) the presence of collimated energy beams  
on the kiloparsec-scale, even though the beams are not seen directly.
These findings further exemplify that non-detection of jets (a possible cause 
of jet asymmetry) could well be a mere illusion (however, see the recent 
detailed optical/Near-IR/UV imaging study of the M87 jet by 
Meisenheimer, R\"oser \& Schl\"otelburg 1996; also, van 
Groningen, Miley, \& Norman 1980). 
\\

\noindent{\s 7.4. Invisible beams: Doppler-hidden, or simply non-existent?}
\\

\noindent
Two oft-cited, non-statistical evidences for the existence of `invisible'
jets are:

(i)   The discovery of a nonthermal optical synchrotron hot-spot at the
expected end-point of the (invisible) counter-jet in M87 (Stiavelli
 et al.\ 1992; Sparks et al.\ 1992).  In that the lifetimes of the 
optically radiating electrons are exceedingly  short ($\le 100~$yr; 
Meisenheimer et al.\ 1996), an active counter-jet has been inferred. 

(ii)   From long-slit spectroscopy of the region of the (invisible) counter-jet
in the superluminal galaxy 3C 120, Axon et al.\ (1989) have found 
evidence for a `cocoon' of thermal gas which contains signatures of
kinematic disturbance arising from the passage of the  putative counter-jet 
through the cocoon.
\\

\noindent{\s 7.5. Does asymmetric dissipation cause apparent jet asymmetry?}
\\

\noindent
Numerous authors have advocated the possibility that the apparent asymmetry
of jets, instead of necessarily arising from relativistic beaming of 
the approaching jet, could even be caused through an enhanced dissipation 
of one of the two jets. Conceivably, this could occur in cases where the 
host galaxy has an asymmetric
environment, leading to a more rapid entrainment of the ambient thermal 
plasma into the relativistic jet on that side (e.g.,  Bicknell 1984).   
It is of considerable interest to examine this possibility for the
origin of jet asymmetry through jet dissipation, particularly for the 
case of powerful one-sided jets observed in quasars. The dissipation
mechanism is also commonly invoked to explain the
absence of hot-spots in low power (FR I) radio galaxies in which the
presence of two-sided jets at large distance from the nucleus is interpreted as
a manifestation of the jet's deceleration through the postulated entrainment
process (e.g., Bicknell 1994; Komissarov 1994; Laing 1995). 

A second possibility 
proposed for the lack of hot-spots invokes a slowing down of the jet's front 
to trans-sonic velocity, leading to its decollimation and the consequent
weakening of the Mach disk (e.g., Gopal-Krishna \& Wiita 1988;
Gopal-Krishna 1991; Roland, Lehoucq \& Pelletier 1992; Blandford 1996).  
Thus, the relativistic jet's
plasma is henceforth discharged into a non-relativistic, plume-like outflow,
rather than into a well confined terminal hot-spot.  We have shown that
this transition may be important in converting FR II sources into FR I's
(Gopal-Krishna \& Wiita 2001).

To  test these alternatives we (Gopal-Krishna, Wiita, \& Hooda 1996)
have examined a sample of 8 `weak-headed-quasars'
(WHQs) which seem to form a significant fraction $(\sim 10\%$) of a 
representative sample of highly luminous radio quasars located at
high-redshifts ($z \ge 1.5$) and
imaged with the VLA at $\lambda = 6~$cm (Lonsdale, Barthel \& Miley 1993). 
The peculiarity marking these 8 sources is that their (one-sided) jets are very
prominent but do not terminate in a conspicuous hot-spot.
On phenomenological grounds we argued that the observed lack of two-sided 
jets among the WHQs seems to be more consistent with the second alternative
mentioned above (Gopal-Krishna et al.\ 1996).  For, if the
entrainment/dissipation model were true, the jet's visibility would not be
determined solely by beaming effects and therefore roughly half of the 8 jets
seen should be on the far side of the nucleus and still be visible.  But
in none of the 8 cases is the putative approaching counterpart of any
such misdirected jet detected as well (even though the radiation 
from any such counterpart would 
be Doppler boosted toward us). The non-detection of two-sided jets in any of the 
WHQs supports the Mach-disk decay
picture; we argued that various other ways of explaining the result
(e.g., hot-spot emission beamed away from us; hot-spot having a much
steeper spectral index than the jet; very asymmetric ambient gas
distributions) are all quite implausible. 

Under the beaming hypothesis, the
jetted side is always the one nearer to us, so it is seen at a later
stage in the source evolution. This time lag would facilitate the
viewing of the shock-front's disruption and weakening of the hot-spot
on the jet's side.  Non-relativistic numerical simulations show that
the Mach-disk decay and hot-spot fading can occur, but the more
appropriate relativistic simulations have yet to be performed.  
Additional good quality maps of high-$z$ quasars should yield a larger
sample of WHQs, which would permit a more persuasive evaluation of the
various competing mechanisms proposed.
\\
\newpage
\noindent{\s 8. Conclusions}
\\

\noindent
Over the past three decades astronomers have successfully wielded the radio source asymmetries 
(or, the lack thereof) as a double-edged knife to attack the twin issues of radio source formation
and evolution. It would be too ambitious on our part to attempt a survey of the enormous 
extant literature on this subject, let alone review it in totality. Therefore we have 
endeavored here to focus on just a few selected facets of the problem, which we hope serve 
to highlight the complex interplay between the intrinsic, environmental
and viewing-dependent factors, leading to the various manifestations of the asymmetries
reported in the literature. 

Independent analyses/interpretations of the observed asymmetries
in the lengths, luminosities, spectra, spectral curvature, depolarization, optical/UV
surface brightness and non-colinearity of the twin-lobes, have provided a number of key
leads and several important conclusions, especially when the relativistic speed of the 
jets is taken into account. As discussed in the previous sections, these findings form 
a crucial part of the growing body of evidence in favor of the orientation-based unification 
theories.  The outcome of the asymmetry studies now also needs to be harnessed for 
exploring the evolutionary connections between the two main morphological classes of 
radio galaxies (FR I and FR II), and for probing the relationships between the dimensional
sequence represented by compact-symmetric objects (CSOs), Gigahertz-peaked-spectrum (GPS)
sources, compact-steep-spectrum (CSS) sources, intermediate-symmetric-objects (ISO),
classical RGs, and giant radio galaxies (GRG) (e.g., Snellen et al. 2000).  

Investigations of radio source asymmetries have raised several outstanding questions 
whose resolution will require continued confrontation of thought with observations.
We sum up this review with the proverbial set of 10 posers related to this topic, 
though the careful reader may note that we have squeezed in several extra follow-up
questions:

\noindent$\bullet$ {Are the mild hot-spot speeds ($\approx c/100$), as indicated by 
the $Q$-distributions of
radio-luminous quasars, consistent with the markedly stronger asymmetry observed for quasars 
(compared to RGs), if orientation is the only difference between the two?}

\noindent$\bullet$ {Do powerful radio sources at high redshifts have more distorted 
radio structures, compared to their less distant counterparts?}

\noindent$\bullet$ {Typically, how large is the `backflow' speed in classical 
double radio sources? What is its dependence on the speed of the hot-spot?}

\noindent$\bullet$ For FR II double radio sources, how reliably can the apparent 
asymmetry of the Ly-$\alpha$ halo be used to identify the near-side lobe? 
How does the Ly-$\alpha$ asymmetry correlate with the lobe depolarization asymmetry 
(which is a fairly robust indicator of the sidedness)?

\noindent$\bullet$ How different are the spectral-index asymmetries for the radio lobes and the hot-spots?
Such information may provide vital clues for disentangling  intrinsic, environmental and
orientational effects on the apparent spectral-index asymmetry.

\noindent$\bullet$ How significant is the contribution of the parent galaxy's 
motion to the lobe-length asymmetry?

\noindent$\bullet$ Will the correlated radio-optical asymmetry turn out to be a weaker effect if red 
emission-line images of the EELR are used for quantifying the optical asymmetry? Such a 
finding would support the notion that the optical asymmetry is primarily a result of 
asymmetric dust obscuration (rather than due to an asymmetric line-emission). 

\noindent$\bullet$ How genuine is the `mystique' of 3C273? Can nature make truly
one-sided jets?


\noindent$\bullet$ Are there RGs (located in extremely rarefied
environments) whose jets are still boring their way at relativistic speed
on the kiloparsec scale (and consequently, have not yet formed 
their hot-spots and lobes)?

\noindent$\bullet$ What is the late phase of the evolutionary scenario for jets? 
What makes their
hot-spots finally decay  (e.g., weak-headed-quasars: WHQs)? Does their decay 
finally 
destabilize the jets themselves? This question also has a strong bearing on 
the issue of 
unification of the two Fanaroff-Riley morphological classes. 

\vspace{0.3cm}
\noindent
{\s Acknowledgment}: We offer our apologies to all those authors whose
work was not cited in this article, in spite of its relevance; these omissions are
not merely due to our ignorance, but also arose from the combination of
limited space and
the vastness of the literature on this topic. We thank J.P.\ Leahy and H.\ 
R\"ottgering for providing the electronic versions of Figs.\ 1 and 2, 
respectively.  This work was supported in part by NASA grant NAG5-3098 and Research
Program Enhancement
funds to the Program in Extragalactic Astronomy at Georgia State University.  PJW
is grateful for continuing hospitality at the Department of Astrophysical Sciences at
Princeton University. 
}

%
\vspace{0.5cm}
\noindent
{\s References}
\\
                                                                                
{\fc

\noindent  Antonucci, R., \& Barvainis, R. 1990, ApJ, 363, L17

\noindent Arshakian, T.G., \& Longair, M.S. 2000, MNRAS, 311, 846

\noindent  Axon, D.J., Unger, S.W., Pedlar, A., Meurs, E.J.A., Whittle, D.M., \& Ward, M.J. 1989, Nature, 341, 631 

\noindent  Baade, W. 1956, ApJ, 123, 550 

\noindent  Banhatti, D.G. 1980, A\&A, 84, 112

\noindent  Barger, A.J., et al. 2001, AJ, 122, 2177

\noindent  Barthel, P.D. 1989, ApJ, 336, 606



\noindent  Barthel, P.D., \& Miley, G.K. 1988, Nature, 333, 319

\noindent  Baryshev, Yu., \& Teerikorpi, P. 1995, A\&A, 295, 11

\noindent  Begelman, M.C., Blandford, R.D., \& Rees, M.J. 1980, Nature, 287, 307

\noindent  Begelman, M.C., Blandford, R.D., \& Rees, M.J. 1984,
Rev.\ Mod.\ Phys.\ 56, 255

\noindent  Best, P.N., Bailer, D.M., Longair, M.S., \& Riley, J.M. 1995, MNRAS, 275, 1171

\noindent  Bicknell, G. 1984, ApJ, 286, 68

\noindent  Bicknell, G. 1994, ApJ, 422, 542

\noindent  Bicknell, G. 1995, ApJS, 101, 29

\noindent  Bisnovatyi-Kogan, G.S. \& Moisenko, S.G. 1992, Astron.\ Zh., 36, 285

\noindent  Black, A.R.S., Baum, S.A., Leahy, J.P., Perley, R.A.,
Riley, J.M., \& Scheuer, P.A.G. 1992, MNRAS, 256, 186

\bpar 
Blandford, R.D. 1996, in Cygnus A -- Study of a Radio Galaxy, C.\ Carilli
\& D.\ Harris, Cambridge: Cambridge Univ. Press, 264

\noindent  Blandford, R.D., \& Rees, M.J. 1974, MNRAS, 169, 395


\noindent  Blundell, K.M., \& Alexander, P. 1994, MNRAS, 267, 241

\noindent  Blundell, K.M., \& Rawlings, S. 2000, AJ, 119, 1111

\bpar  
Blundell, K.M., \& Rawlings, S. 2001, in Particles and Fields in Radio Galaxies,
ASP Conf.\ Ser.\ 250, R.A.\ Laing \& K.M.\ Blundell, San Francisco: ASP, 363

\noindent  Blundell, K.M.,  Rawlings, S., \& Willott, C.J. 1999, AJ, 117, 677

\bpar
Bodo, G., Chagelishvili, G.D., Ferrari, A., Lominadze, J.G., \& Trussoni, E. 1990, in Plasma Astrophysics, ESA SP-311, T.D.\ Guyenne \& J.J. Hune, Paris: ESA, 273

\noindent  Bregman, J.N., Hogg, D.E., \& Roberts, M.S.  1995, ApJ, 441, 561 

\bpar  
Bridle, A.H. 1996,  in Energy Transport in Radio Galaxies and Quasars, ASP Conf.\ Ser.\ 100, 
 P.E.\ Hardee, A.H.\ Bridle, \&
J.A.\ Zensus, San Francisco: ASP, 383


\noindent  Bridle, A.H., \& Perley, R.A. 1984, ARA\&A, 22, 319

\bpar
Bridle, A.H., Laing, R.A., Scheuer, P.A.G. \& Turner, S. 1994, in The Physics of Active Galaxies, G.V.\ Bicknell, M.\ Dopita, P.\ Quinn, ASP Conf.\ Ser.\ 54, San Francisco: ASP, 187

\noindent  Browne, I.W.A., \& Perley, R.A. 1986, MNRAS, 222, 149

\noindent  Capetti, A., Morganti, R., Parma, P., \& Fanti, R. 1993, A\&AS, 99, 407

\noindent  Chambers, K.G., Miley, G.K., van Breugel, W.J.M.  1990, ApJ, 363, 21 


\noindent  Carilli, C. L., Owen, F. N. \& Harris, D. E. 1994, AJ, 107, 480

\noindent Cen, R. \& Ostriker, J. 1999, ApJ, 519, L109

\noindent Chiaberge, M. 2004, astro-ph/0405255

\noindent  Chini, R., \& Kr\"ugel, E.  1994, A\&A, 288, L33

\noindent  Chy\.zy, K.T., \& Zi{\c e}ba, S. 1995, A\&A, 303, 420 

\noindent  Cimatti, A., di Serego Alighieri, S., Fosbury, R.A.E., Salvati, M., \& Taylor, D. 1993,  MNRAS, 264, 421

\bpar 
Clarke, D.A. 1996, in Energy Transport in Radio Galaxies and Quasars, ASP Conf.\ Ser.\ 100,  P.E.\ Hardee, A.H.\ Bridle, \&
J.A.\ Zensus, San Francisco: ASP, 311

\noindent  Cohen, M.H. \& Kellermann, K.I. 1995, Proc.\ Natl.\ Acad.\ Sci.\ USA, 92, 11339

\noindent  Colina, L., \& de Juan, L. 1995, ApJ, 448, 548

\noindent  Conway, R.G., Davis, R.J., Foley, A.R., \& Ray, T.P.  1981, Nature, 294, 540


\noindent  Cowie, L.L.,  Hu, E.M.,  \& Songaila, A. 1996, Nature, 377, 603

\noindent  Curtis, H.D. 1918, Lick Obs.\ Publ., 13, 11


\noindent  Dav{\'e}, R., et al. 2001, ApJ, 355, 416

\noindent  Davis, R.J., Muxlow, T.W.B., \& Conway, R.G. 1985, Nature, 318, 343

\noindent  Dennett-Thorpe, J., Bridle, A.H.,  Scheuer, P.A.G. Laing, R.A., Leahy, J.P. 1997,
MNRAS, 289, 753


\noindent  De Young, D.S., \& Axford, W.I. 1967, Nature, 216, 129

\noindent  Dunlop, J.S., Hughes, D.H., Robbins, S., Eales, S.A., \& Ward, M.J.
1994, Nature, 370, 347

\noindent  Eilek, J.A., \& Arendt, P.N. 1996, ApJ, 457, 150

\noindent  Ensman, L.M., \& Ulvestad, J.S. 1984, AJ, 89, 1275

\noindent  Evrard, A.E., Summers, F.J., \& Davis, M. 1994, ApJ, 422, 11 

\noindent  Fokker, A.D. 1986, A\&A, 156, 315

\noindent  Fomalont, E.B. 1969, ApJ, 157, 1027

\noindent  Fraix-Burnet, D. 1992, A\&A, 259, 445

\noindent  Fukugita, M., Hogan, C.J.,  \& Peebles, P.J.E. 1996, Nature, 381, 489 

\noindent  Garrington S.T., \& Conway, R.G. 1991, MNRAS 259, 198

\noindent  Garrington S.T., Conway, R.G., \& Leahy, J.P. 1991, MNRAS, 250, 171

\noindent  Garrington, S.T., Leahy, J.P., Conway, R.G., \& Laing, R.A. 1988, Nature, 331, 147

\noindent  Georganopoulos, M., \& Kazanas, D. 2003, ApJ, 589, L5

\noindent  Giraud, E., Melnick, J., Gopal-Krishna, Oliveira, C. \& Kulkarni, V.K. 1996,
 A\&A, 309, 733

\noindent  Gopal-Krishna 1980, A\&A, 86, L1

\noindent  Gopal-Krishna 1991, A\&A, 248, 415

\noindent  Gopal-Krishna 1995, Proc.\ Nat.\ Acad.\ Sci.\ USA, 92, 11399


\noindent  Gopal-Krishna, Giraud, E., Melnick, J., \& Della Valle, M.  1995, A\&A, 303, 705 

\noindent  Gopal-Krishna, Kulkarni, V.K., \& Wiita, P.J. 1996, ApJ, 463, L1

\noindent  Gopal-Krishna \& Nath, B. 1997, A\&A, 326, 45

\noindent  Gopal-Krishna \& Wiita, P.J. 1988, Nature, 333, 49

\noindent  Gopal-Krishna \& Wiita, P.J. 1996, ApJ, 467, 191 (GW96)

\noindent  Gopal-Krishna \& Wiita, P.J. 2000, ApJ, 529, 189 (GW00)

\noindent  Gopal-Krishna \& Wiita, P.J. 2001, A\&A, 373, 100

\noindent  Gopal-Krishna,  Wiita, P.J., \& Hooda, J.S. 1996, A\&A, 316, L13

\noindent  Gopal-Krishna,  Wiita, P.J., \& Saripalli, L. 1989, MNRAS, 239, 173

\noindent  Goudfrooij, P. \& de Jong, T. 1995, A\&A, 298, 784 

\noindent  Hargrave, P.J., \& McEllin, M. 1975, MNRAS, 173, 37 

\noindent  Hargrave, P.J., \& Ryle, M. 1974, MNRAS, 166, 305

\noindent  Heckman, T.M., Lehnert, M.D., van Breugel, W., \& Miley, G.K.  1991a, ApJ, 370, 78 

\noindent  Heckman, T.M., Lehnert, M.D., Miley, G.K., \& van Breugel, W. 1991b,
ApJ, 381, 373

\noindent  Hintzen, P., \& Scott, J.S. 1978, ApJ, 224, L47

\noindent  Hippelein, H. \& Meisenheimer, K. 1993,  Nature, 362, 224 

\noindent Hooda, J.S., \& Wiita, P.J. 1998, ApJ, 493, 81

\noindent  Hough, D.H. \& Readhead, A.C.S. 1989, AJ, 98, 1208

\noindent  Icke, V. 1983, ApJ, 265, 648

\noindent  Ingham, W., \& Morrison, P. 1975, MNRAS, 173, 569

\noindent  Ishwara-Chandra, C.H., \& Saikia, D.J. 2002, New Astr.\ Rev., 46, 71

\noindent  Jennison, R.C., \& Das Gupta, M.K. 1953, Nature, 172, 996

\noindent  Jeyakumar, S., Wiita, P.J., Saikia, D.J., \& Hooda, J.S., 2004, A\&A, in press

\noindent  Johnson, R.A., Leahy, J.P., \& Garrington, S.T. 1995, MNRAS, 273, 877

\noindent  Kapahi, V.K. 1989, AJ, 97, 1

\bpar
Kapahi, V.K. 1990a, in Parsec-scale Radio Jets, J.A.\ Zensus 
\& T.J.\ Pearson, Cambridge: Cambridge University Press, 304

\noindent  Kapahi, V.K. 1990b, Current Sci., 59, 561

\noindent  Kapahi, V.K., \& Saikia, D.J. 1982, J.\ Ap.\ Astr., 3, 465

\noindent  Kellermann, K.I. et al. 2004, ApJ, 609, 539

\noindent  Komberg, B.V. 1994, Astr.\ Rep., 38, 617

\noindent  Kommissarov, S.S. 1994, MNRAS, 269, 394

\noindent  Kotanyi, C.G.,  \& Ekers, R.D. 1979, A\&A, 73, L1

\noindent  Kundt, W., \& Gopal-Krishna, 1980, Nature, 288, 149 

\noindent  Kundt, W., \& Gopal-Krishna, 1981, Ap.\ Space Sci., 75, 257

\noindent  Kundt, W., \& Gopal-Krishna, 1986, J.\ Ap.\ Astr., 7, 225

\noindent  Lacy, M., et al. 1994, MNRAS, 271, 504 

\noindent  Laing, R.A., 1988, Nature, 331, 149


\noindent  Laing, R.A. 1995, Proc.\ Natl.\ Acad.\ Sci.\ USA, 92, 11413

\noindent  Lawrence, A. 1991, MNRAS, 252, 586

\noindent  Leahy, J.P., Muxlow, T.W.B., \& Stephens, P.W. 1989, MNRAS, 239, 401 

\noindent  Leahy, J.P., \& Perley, R.A. 1991, AJ, 102, 537

\noindent  Lister, M.L., Gower, A.C., \& Hutchings, J.B. 1994, AJ, 108, 821

\noindent  Lister, M.L., Hutchings, J.B., \& Gower, A.C. 1994, ApJ, 427, 125 (LHG)

\noindent  Liu, R. \& Pooley G.G. 1991a, MNRAS, 249, 343

\noindent  Liu, R. \& Pooley G.G. 1991b, MNRAS, 253, 669

\noindent  Liu, R., Pooley, G.G., \& Riley, J.M. 1992, MNRAS, 257, 545


\noindent  Longair, M.S., \& Riley, J.M. 1979, MNRAS, 188, 625

\noindent  Longair, M.S., Ryle, M., \& Scheuer, P.A.G. 1973, MNRAS, 164, 243

\noindent  Lonsdale, C.J., Barthel, P.D., \& Miley, G.K. 1993, ApJS, 87, 63


\noindent  Lynden-Bell, D. 1996,  MNRAS, 279, 389

\noindent  Macdonald, G.H., \& Miley, G.K. 1971, ApJ, 164, 237 

\noindent  Mackay, C.D. 1971, MNRAS, 154, 209

\noindent  Mackay, C.D. 1973, MNRAS, 162, 1

\noindent  Macklin, J.T. 1981, MNRAS, 196, 967

\noindent  Mangalam, A.V. \& Gopal-Krishna 1995, MNRAS 275, 979

\noindent  Marconi, A., Risaliti, G., Gilli, R., Hunt, L.K., Maiolino, R., Salvati, M.
2004, MNRAS, 351, 169

\noindent  Matveyenko, L.I. et al. 1980, SvAL, 6, 42

\bpar 
Matveyenko, L.I., Pauliny-Toth, I.I.K., Graham, D.A., Sherwood, W.A., 
B\"aath, L.B., \& Kus, A.D. 1996, Astr.\ Lett., 22, 14

\bpar
McCarthy, P.J., Spinrad, H., van Breugel, W., Liebert, J., Dickinson, M., Djorgovski, \& Eisenhardt, P. 1990, ApJ, 365, 487

\bpar
McCarthy, P.J. \& van Breugel, W. 1989, in Extranuclear Activity in Galaxies, E.J.A.\ Meurs \& R.A.E.\ Fosbury, Garching: ESO, 55

\bpar
McCarthy, P.J., van Breugel, W., \& Kapahi, V.K. 1991, ApJ, 371, 478 (MvBK)

\noindent  Meisenheimer, K. \& Hippelein, H. 1992, A\&A, 264, 465 


\noindent  Meisenheimer, K., R\"oser, H.-J. \& Schl\"otelburg, M. 1996, A\&A,
307, 61

\noindent  Mellema, G., Kurk, J.D., \& R{\"o}ttgering, H.J.A. 2002, A\&A, 395, L13


\noindent  Miley, G.K. 1980, ARA\&A, 18, 165 

\noindent  Miller, C.J., Nichol, R.C., G{\'o}mez, P.L., Hopkins, A.M., \& 
Bernardi, M. 2003, ApJ, 597, 142

\noindent  Morisawa, K. \& Takahara, F.  1987, MNRAS, 228, 745

\noindent  Morrison, P., Roberts, D., \& Sadun, A. 1984, ApJ, 280, 483

\noindent  Morrison, P., \& Sadun, A. 1992, MNRAS, 254, 488

\bpar
Norman, M.L. 1996, in Energy Transport in Radio Galaxies and Quasars, ASP Conf.\ Ser.\ 100,  P.E.\ Hardee, A.H.\ Bridle, \&
J.A.\ Zensus, San Francisco: ASP, 319


\noindent  Parma, P., de Ruiter, H.R., Mack, K.-H., van Breugel, W., Dey, A., Fanti, R., \& Klein, U. 1994, A\&A, 311, 49

\noindent  Peacock, J.A. 1987, in Astrophysical Jets and their Engines,  W.\ Kundt,     Dordrecht: Reidel, 185

\noindent  Pedelty, J.A., Rudnick, L., McCarthy, P.J., \& Spinrad, H.  1989, AJ, 97, 647




\noindent  Rees, M.J. 1988, MNRAS, 239, 1P

\noindent  Rees, M.J., Begelman, M. C., Blandford, R. D. \& Phinney, E. S. 1982,
Nature, 295, 17

\bpar
Riley, J.M \& Jenkins, C.J. 1977, in Radio Astronomy and Cosmology, IAU Symp.\ 74,  D.L. Jauncey, Dordrecht: Reidel, 237


\bpar
Roland, J., Lehoucq, R., \& Pelletier, G. 1992, in Extragalactic Radio Sources--from Beams to Jets, J.\ Roland, H.\ Sol, G.\ Pelletier, Cambridge: Cambridge University Press, 294

\bpar
Romney, J.D., Biretta, J.M., Dhawan, V., Kellermann, K.I., Vermeulen, R.C., \& Walker, R.C. 1995, Proc.\ Natl.\ Acad.\ Sci.\ USA, 92, 11360

\noindent  Rudnick, L., \& Edgar, B.K. 1984, ApJ, 279, 74

\noindent  Rudnick, L., Katz-Stone, D.M., \& Anderson, M.C. 1994, ApJS, 90, 955

\noindent  Ryle, M., \& Longair, M.S. 1967, MNRAS, 136, 123

\noindent  Ry\'s, S. 1994, A\&A, 281, 15

\noindent  Saikia, D.J. 1984, MNRAS, 245, 408

\noindent  Saikia, D.J., Junor, W., Cornwell, T.J., Muxlow, T.W.B., \& Shastri, P. 1990, MNRAS, 245, 408

\noindent  Saikia, D.J., Jeyakumar, S.K., Wiita, P.J., Sanghera, H.S., \& Spencer, R.E. 1995, MNRAS, 276, 1215

\bpar
Saikia, D.J., Jeyakumar, S.K., Mantovani, F., Salter, C.J., Spencer, R.E., Thomasson, P.,
\& Wiita, P.J. 2003, PASA, 20, 50

\noindent  Saikia, D.J., \& Wiita, P.J. 1982, MNRAS, 200, 83

\noindent  Saripalli, L, Gopal-Krishna, Reich, W., \& K\"uhr, H. 1986, A\&A, 170, 20

\noindent  Scheuer, P.A.G. 1974, MNRAS, 166, 513

\bpar
Scheuer, P.A.G. 1982, in Exragalactic Radio Sources, IAU Symp.\ 97, D.S.\ Heeschen, C.M.\ Wade, Dordrecht: Reidel, 163

\noindent  Scheuer, P.A.G. 1987, in Astrophysical Jets and their Engines,  W.\ Kundt,     Dordrecht: Reidel, 129

\noindent  Scheuer, P.A.G. 1995, MNRAS, 277, 331



\noindent  Siah, J., \& Wiita, P.J. 1990, ApJ, 363, 411

\noindent  Singal, A.K. 1988, MNRAS, 233, 87

\noindent  Slysh, V.I. 1966, Sov.\ Astr., 9, 533

\bpar
Snellen, I.A.G., Schilizzi, R.T., Miley, G.K., de Bruyn, A.G.,
Bremer, M.N. \& R\"ottgering, H.J.A. 2000, MNRAS, 319, 445

\noindent  Sofue, Y., \& Wakamatsu, K. 1992, PASJ, 44, L23

\noindent  Sparks, W.B., Fraix-Burnet, D., Macchetto, F., \& Owen, F.N. 1992, Nature, 355, 804

\noindent  Spinrad, H., Dey, A., \& Graham, J.R.  1995, ApJ, 438, L51

\noindent  Stawarz, L. 2004, ApJ, 613, 119

\noindent  Stiaveli, M., Biretta, J., M\/oller, P., Zeilinger, W.W.  1992, Nature, 355, 802

\noindent  Strom, R.G. \& J\"agers, G. 1988, A\&A, 194, 79

\noindent  Swarup, G., \& Banhatti, D.G. 1981, MNRAS, 194, 1025

\noindent  Teerikorpi, P. 1984, A\&A, 132, 179

\noindent  Teerikorpi, P. 1986, A\&A, 164, L11

\noindent  Tribble, P.C.  1992, MNRAS, 256, 281

\noindent  Urry, C.M., \& Padovani, P. 1995, PASP, 107, 803

\noindent  Valtonen, M.J. 1979, ApJ, 231, 312

\noindent  van Groningen, E., Miley, G.K., \& Norman, C. 1980, A\&A, 90, L7 

\bpar
van Ojik, R., R\"ottgering, H.J.A., Carilli, C.L., Miley, G.K.,
Bremer, M.N. \& Macchetto, F. 1996, A\&A, 313, 25


\noindent  Vermeulen, R.C., Readhead, A.C.S., \& Backer, D.C. 1994, ApJ, 429, L41 

\noindent  Wang, J.C.L., Sulkanen, M.E., \& Lovelace, R.V.E. 1992, ApJ, 390, 46

\noindent  Wang, Z.,  Wiita, P.J., \& Hooda, J.S. 2000, ApJ, 534, 201

\noindent  Wiita, P.J. 1978, ApJ, 221, 41

\bpar
Wiita, P.J. 1990, in Beams and Jets in Astrophysics,  P.A.\ Hughes, Cambridge: Cambridge University Press,  379

\bpar
Wiita, P.J. 1996, in Energy Transport in Radio Galaxies and Quasars, ASP Conf.\ Ser.\ 100,  P.E.\ Hardee, A.H.\ Bridle, \&
J.A.\ Zensus, San Francisco: ASP, 395

\noindent  Wiita, P.J., \& Gopal-Krishna 1990,
ApJ, 353, 476

\noindent  Wiita, P.J., \& Siah, M.J. 1981, ApJ, 243, 710


\noindent  Willis, A.G., Wilson, A.S., \& Strom, A.G. 1978, A\&A, 66, L1

\bpar
Wills, B.J., Netzer, H., Brotherton, M.S., Han, M., Wills, D., Baldwin,
J.A., Ferland, G.J., \& Browne, I.W.A.  1993, ApJ, 410, 534

\noindent  Wilson, A.S. \& Colbert, E.J.M. 1995, ApJ, 438, 62 


\noindent  Yee, H.K.C., \& Ellingson, E.  1993, ApJ, 411, 43

\noindent York, D. G. et al. 2000, AJ, 120, 1579

\noindent  Zi{\c e}ba, S. \& Chy\.zy, K.T. 1991, A\&A, 241, 22

}

                                                                                



\end{document}